\documentclass[preprint,10pt]{aastex}
\usepackage{psfig,dynamics}

\begin{document}

\title{On the Early Evolution of Forming Jovian Planets II:
Analysis of Accretion and Gravitational Torques}
\author{Andrew F. Nelson\footnote{UKAFF Fellow. Present address: School
of Mathematics, University of Edinburgh, Edinburgh Scotland
EH9~3JZ} }
\affil{Max Planck Institut f\"ur Astronomie, K\"onigstuhl 17, D-69117
Heidelberg, Germany}
\author{Willy Benz}
\affil{Physikalisches Institut, Universit\"at Bern, Sidlerstrasse 5, 
CH-3012 Bern, Switzerland}

\begin{abstract}

We continue our numerical study of the migration of an already formed
proto-Jovian companion embedded in a circumstellar disk. We first
study the sensitivity of the planet's migration to its mass accretion
rate, and find that the disk can supply a forming planet with mass at
an essentially infinite rate ($\sim1$\mj/25~yr) so that a gap could
form very quickly via further dynamical interactions between the
planet and remaining disk matter. The accreted matter has less orbital
angular momentum than the planet and exerts an effective inward
torque, so that inward migration is slightly accelerated. However, if a
partial gap is formed prior to rapid accretion, the effective torque
is small and its contribution to the migration is negligible. Although
the disk can supply mass at a high rate, we show that mass accretion
rates faster than $\sim10^{-4}$\mj/yr are not physically reasonable in
the limit of either a thin, circumplanetary disk or of a spherical
envelope. Planet growth and ultimately survival are therefore limited
to the planet's ability to accept additional matter, not by the disk
in which it resides.

Large gravitational torques are produced both at Lindblad resonances
and at corotation resonances. We compare the torques in our
simulations to analytic theories at Lindblad resonances and find that
common approximations to the theories predict torques that are a
factor $\sim10$ or more larger than those obtained from the
simulations. Accounting for the disk's vertical structure (crudely
modeled in our simulations and the theory with a gravitational
softening parameter) and small shifts in resonance positions due to
pressure gradients, to disk self gravity and to inclusion of non-WKB
terms in the analysis \citep{art93a} can reduce the difference to a
factor $\sim3-6$, but do not account for the full discrepancy. Torques
from the corotation resonances that are positive in sign, slowing the
migration, contribute 20-30\% or more of the net torque on the planet,
but are not well resolved and vary from simulation to simulation. A
more precise accounting of the three dimensional mass distribution and
flow pattern near the planet will be required to accurately specify
the torques from both types of resonances in the simulations. 

We show that the assumption of linearity underlying theoretical
analyses of the interactions at Lindblad resonances is recovered in
the simulations with planets with masses below 0.5\mj, but the
assumption that interactions occur only at the resonances may be more
difficult to support. Angular momentum transfer occurs over a region
of finite width near both Lindblad and corotation resonances. The
shape of the disk's response there (due e.g. to local variations in
epicyclic frequency) varies from pattern to pattern, making the true
position of the resonance less clear. We speculate that the finite
width allows for overlap and mixing between resonances and may be
responsible for the remainder of the differences between torques from
theory and simulation, but whether accounting for such overlap in a
theory will improve the agreement with the simulations is not clear.

\end{abstract}


In a companion paper \citep[][hereafter \jone]{Jov1}, we began a study
of the early stages of the migration of a Jovian planet through an
accretion disk, using numerical simulations. That work was a study of
several important systematic effects that affect the results of
numerical simulations and an exploration of a very large parameter
space of systems. The parameter study was designed to explore the
range of possible systems likely to be encountered during the
evolution of real circumstellar disks and to make a number of
qualitative comparisons of our results to theory. In this work, we
concentrate our study on two of the outstanding issues raised in that
work. 

The first issue concerns our finding that a planet could not complete
a transition to \ttwo\  migration, unless it was more massive than
0.3\mj, and that the migration rate was both very rapid and very
sensitive to the details of the mass distribution within 1--2Hill
radii of the planet. The second issue concerns our finding that the
migration rates did not follow the theoretically expected linear
dependence on planet mass, but that the \mplan$^{2/3}$ proportionality
predicted for the gap width and based on the same physical model did
hold.

We study these two issues through a detailed investigation of mass
accretion and of the gravitational torques between the disk and
planet, and their origin. In section \ref{sec:initconds}, we outline
the initial conditions for an additional series of simulations, not
discussed in \jone (the `{\it gap}' series of simulations), how we
implement mass accretion in our code and the physical significance we
can derive from it. We also summarize the main theoretical results for
the derivations of gravitational torques between the planet and disk.
In section \ref{sec:acc-import}, we look at accretion onto the planet,
whether it can affect the migration rate, and given a lower limit on
the accretion rate we discuss the likely morphology of the object. In
section \ref{sec:interact}, we will examine in detail four simulations
from \jone\  (the low mass and high resolution prototypes with and
without disk self gravity) in order to explore in quantitative detail
the differences between our hydrodynamical simulations and linear
analyses. Our purpose will be to determine the extent to which linear
theory may be applied to the problem, as well as where and how it may
fail. Lastly, in section \ref{sec:summary}, we compare our results to
others in the literature.

\section{The initial conditions and input physics}\label{sec:initconds}

In this work we will make use of the `low mass' and `high resolution'
prototypes, discussed in detail in \jone. Briefly summarized, these
models consists of a circumstellar disk with mass \mdisk$=$0.05\msun,
that extends from 0.5 to 20~AU. Mass is distributed according to an
$r^{-3/2}$ power law and the temperature is defined by an $r^{-1/2}$
power law and the condition that the temperature at 1~AU is $T=250$~K.
We use an isothermal equation of state and the boundary conditions at
the inner and outer edges of the grid are reflecting. For the low mass
prototype, the disk is resolved on a 128$\times$224
cylindrical\footnote{Throughout this paper and its companion, \jone,
we use `$\theta$' to denote the azimuth coordinate rather than the
more usual variable `$\phi$' in order to avoid confusion between
references to the coordinate and to components of the planet's
gravitational potential, $\phi_m$, common throughout this work.} grid
$(r\times\theta)$, while for the high resolution prototype we doubled
the grid resolution in each dimension, to $256\times448$.

The planet is set into an initially circular orbit at 5.2~AU and is
thereafter influenced by gravitational forces from the 1\msun\  central
star (fixed to the origin) and the disk. The gravitational softening
required was defined to be equal to the size of the grid zone where
the planet is found at each time. In order that the simulations would
be identical, the high resolution prototype defined the softening to
be twice the current cell size, so that the absolute values were
identical. We have simulated two versions of these models, one with
and the other without disk self gravity. All of our simulations
studying accretion will include self gravity. 

For our study of accretion we will use initial conditions and a
physical model that are identical to our low mass prototype. We will
perform two separate studies, one beginning from the low mass
prototype, and the other with an already formed gap in the disk. We
produce a gap from the unperturbed initial state by allowing the
system to `pre-evolve' for 3000~yr. In this pre-evolution, we begin
with the low mass prototype, except that we assume that while the
planet exerts gravitational forces on the disk, it is not in turn
affected by the disk. Its orbit therefore remains fixed at 5.2~AU.
During the pre-evolution, spiral patterns are generated by the
planet/disk interaction and a gap structure forms in the disk. We show
the unperturbed profile and the azimuth average of the mass
distribution after 3000~yr in figure \ref{fig:sd-conds}. While it is
clear that the evolution has produced a significant change in the
profile, the gap region is not totally devoid of matter. We monitored
the profile during the evolution and found no significant changes
during the last 1000~yr of this pre-evolution. We therefore believe
that the system is at or near a steady state. 

After the simulation has completed its pre-evolution, we save the
resulting state `as is' and use it as an initial condition for
multiple simulations in which we `turn on' the disk-on-planet gravity
and allow the planet to migrate. One additional modification is
required between the pre-evolution simulation and each of the models
based on it. Turning on this additional gravity term requires that
disk gravity be accounted for in the planet's velocity. We assume
that the its orbit is circular and calculate the angular velocity
required for its orbit from the net gravitational forces due to the
disk and star. 

\subsection{The mechanics of accretion from a grid, and what we can
learn from it}\label{sec:num-acc}

In the {\it acc} and {\it gap} series' of simulations (see definitions
in section \ref{sec:acc-import}) we study mass accretion onto the
forming planet. In these simulations, we allow mass to accrete onto
the planet at a fraction, $f$, of the Bondi-Hoyle-Littleton (BHL) rate
\citep[see e.g.][]{Shu2}:
\begin{equation}\label{eq:BHL-acc}
\dot M_{\rm pl} =   4\pi\rho_\infty f
               {{ (GM_{\rm pl})^2}\over{(V^2 + c_\infty^2)^{3/2}}}.
\end{equation}
We define $c_\infty$ as the sound speed of the gas in the disk at the
orbit radius of the planet and the velocity at infinity, $V$, is taken
as the relative velocity of the gas with respect to the planet at the
Hill radius, \rh$= a_{\rm pl} (M_{\rm pl}/(3M_*))^{1/3}$, averaged
over the azimuth region between the Lagrange points L4 and L5. The
density at infinity is defined by an averaged surface density divided
by twice the scale height, $\rho_\infty=\bar\Sigma/(2H)$. Guided by
the results of \jone, we determine the surface density average using
the radial region between $a_{\rm pl}\pm$2\rh\  and, as for the
velocity, use the azimuthal region between the Lagrange points, L4 and
L5.

At each time step an amount of mass $\dot M_{\rm pl}\delta t$ is added
to the planet and removed from the disk region within one Hill radius
of the planet. The mass removed from each grid zone is computed as
\begin{equation}\label{eq:mdot}
dM_{\rm zone} =  -\dot M_{\rm pl} \delta t 
     \left[2\left(1 - \left({{\mathcal{R}}\over{R_H}}\right)^2\right)\right]
                 {{r_i \delta r_i \delta \theta_j}\over{ \cal A}}
\end{equation}
where $\mathcal{R}$ is the distance of the zone center from the
planet, $r_i\delta r_i\delta\theta_j$ defines the area of a grid zone
and surface area of the Hill sphere, $\cal A$, is taken as the sum
over the areas of all grid zones for which $\mathcal{R}<$\rh. We limit
$dM_{\rm zone}$ to at most half of the mass in the zone, to avoid any
potential instabilities caused by the sudden mass loss. With this
definition, most of the mass removed is taken from the inner half of
the Hill sphere nearest the planet, and no single zone suffers an
unduly large change between successive time steps.

The accretion of mass onto the planet will of course be accompanied by
some amount of angular momentum, both as an orbital component (around
the star) and a spin component. We add this momentum to the planet and
subtract it from each grid zone, so that the net result of the
accretion conserves total angular momentum. The mass accreted onto the
planet will change its orbital angular momentum both by increasing
the planet's mass and by changing its trajectory. Mathematically, the
change in orbital angular momentum of the planet due to accreting
matter can be decomposed into three parts as
\begin{equation}\label{eq:accrete-torq}
{{dJ_z}\over{dt}} = {{d M_{\rm pl}}\over{dt}} rV_\theta
                     + M_{\rm pl}{{d}\over{dt}}\left(rV_\theta\right)
                     - {{dS_z}\over{dt}},
\end{equation}
where $r$ and $V_\theta$ are the planet's orbit radius and azimuthal
velocity. The first term in this equation is due to the accretion of
mass from the disk which is already co-moving with the planet. In
other words, this change in angular momentum (`torque') has no net
effect on the planet's trajectory. The second term represents the
dynamical effect of the mass accretion and can be compared to the
gravitational torques on the planet due to the generation of spiral
density structures in the disk. We discuss the importance of this
effect in section \ref{sec:accrete}. The third term in equation
\ref{eq:accrete-torq} is necessary to properly account for the fact
that the spin of the planet will also be changed by the accretion.
This is because mass accreted onto the planet will in general not have
zero angular momentum, as calculated around the center of mass of the
planet and accreted mass.

While a Bondi-like accretion rate is not strictly valid for the
geometry and flow pattern we consider here, we believe it represents
an interesting limit. It is at or near the highest rate at which
matter can be supplied to the planet by the disk and at or near the
highest rate that the planet can accrete matter because it assumes
that all matter within reach of the planet will be accreted and that
none is able to build up into a long-lived envelope and oppose further
infall. By including accretion or not we will be able to determine
whether mass accreting onto the planet can bring with it enough
orbital angular momentum (an effective torque), to significantly
change the orbit of the planet. We will also be able to explore the
conditions in the circumplanetary environment. Under what conditions
is it reasonable to expect a circumplanetary disk to form? Should we
expect the environment to be a somewhat spherical envelope, or
something still more complex?

\subsection{Analytic torque formulae}\label{sec:torq-analytic}

In section \ref{sec:interact}, we shall make detailed comparisons of
the gravitational torques on the planet as produced in our
simulations, to those predicted by theory. Here we will briefly state
the torque formulae to which we make those comparisons. Analytic
methods typically proceed by linearizing the equations of motion and
evaluating the gravitational torques of the disk and planet on each
other as the sum contributions from a Fourier series of density
perturbations in the disk. The density perturbations are generated at
an inner or outer Lindblad resonance (ILR or OLR, respectively) or a
corotation resonance (CR). 

\citet[][hereafter GT79, GT80]{GT79,GT80} found that by far the
largest portion of the torque exerted by a perturber is transmitted in
the neighborhood of a resonance, and developed approximations for
these torques that were based upon this fact. Because the torques are
transmitted to and from the disk only at resonances, the torque on the
planet can be expressed as the change in angular momentum flux through
the disk at that resonance. Under this approximation, they found that
the angular momentum flux generated by an $m$'th order spiral pattern
at Lindblad resonances is
\begin{equation}\label{eq:LR-flux}
F_{LR}= m\pi^2 f_c \left[\left|{{\Sigma(r)}\over{r dD/dr}}\right|
           \left( r{{d\phi_m}\over{dr}} +
           {{ 2\Omega}\over{\Omega -\Omega_{\rm pl}}}\phi_m 
                          \right)^2 \right]_{r_L}
\end{equation}
where $D$ is the `resonant denominator' and is defined as
\begin{equation}\label{eq:D}
D=\kappa^2-m^2(\Omega -\Omega_{\rm pl})^2.
\end{equation}
$\Omega$ and $\Omega_{\rm pl}$ are respectively, the orbit frequencies
of each location in the disk and of the planet, and square of the
epicyclic frequency is $\kappa^2 = 4\Omega^2 + 2r\Omega{d\Omega/dr}$.

The cutoff function, $f_c$, is introduced artificially to avoid
numerically infinite torques for large $m$ values. Its specific form
is unclear, however \citet{art93a,art93b} has given physical
motivation for it in terms of the modification of the resonance
locations due both to the pressure contribution to the rotation curve
and to averaging the potential of the planet over the vertical
thickness of the disk. He also derives an analytic approximation for
it in the limit of an infinitely thin disk as
\begin{equation}\label{eq:cutoff}
f_c = {{1}\over{H(1+ 4\xi^2)}}
           \left[ {{ 2HK_0(2H/3) + K_1(2H/3)}\over
              {2K_0(2/3) + K1(2/3)}}   \right]^2
\end{equation}
where $K_0$ and $K_1$ are Bessel functions, $H=\sqrt{1 + \xi^2}$ and
$\xi= m(c_s/r\Omega)_p$. 

GT79 also approximate the flux due to interactions near corotation by 
\begin{equation}\label{eq:CR-flux}
F_{CR} = {{- m \pi^2}\over{4}}{\rm sgn}(x)\left[
              \left({\phi^2_m\over{{d\Omega}/dr}}\right)
           {{d}\over{dr}}\left({{\Sigma}\over{B}}\right)
           \right]_{r_c}e^{-|qx|}
\end{equation} 
where $B=\kappa^2/4\Omega$ is the Oort parameter,
$q=(r\kappa/c_s)_{r_c}$, $x=(r-r_{c})/r_{c}$ and $r_c$ is the
corotation resonance location. In both equation \ref{eq:LR-flux} and
\ref{eq:CR-flux}, the gravitational potential of the planet, $\phi_m$,
is one term in an infinite series expansion in Laplace coefficients,
$b^m_{1/2}$. In our comparisons, we calculate the Laplace coefficients
using the method of \citet{Olvers}.

Equations \ref{eq:LR-flux} and \ref{eq:CR-flux} have been further
refined by, e.g. \citet{Ward88,art93a,Ward97} and \citet{Takeuchi96}.
In particular for the LRs, the inclusion of non-WKB terms in the wave
equation, and will result in a modified resonant denominator
\begin{equation}\label{eq:Dstar}
D_* = D + \left({{mc_s}\over{r}}\right)^2, 
\end{equation}
which modifies the effective resonance positions for higher order $m$
patterns by creating a buffer region around the planet, where the
resonances would normally fall. A similar quantity that accounts for
disk self gravity (Artymowicz 2001, personal communication) is
\begin{equation}\label{eq:Dsg}
D_{sg} = D + \left({{mc_s}\over{r}}\right)^2 
                    - 2\pi G\Sigma \left({{m}\over{r}}\right).
\end{equation}
In some analyses, the Laplace coefficients in the potentials are
approximated by Bessel functions,
\begin{equation}\label{eq:Lap-Bes}
b^m_{1/2}(\alpha) \approx {{2}\over{\pi}} K_0(m|1 - \alpha|)
\end{equation}
where $\alpha=r/a_{\rm pl}$, $a_{\rm pl}$ is the semi-major axis of the
planet and $r$ is evaluated at the Lindblad resonances. For an exactly
Keplerian rotation curve, the argument of the Bessel function is
nearly constant (at 2/3) for all $m$ patterns at all Lindblad
resonances. Both \citet{Ward88} and \citet{KP93} have defined a
generalized Laplace coefficient in very similar ways in order to avoid
numerical singularities at the planet's orbit radius. Using the form
from \citet{KP93}, it is
\begin{equation}\label{eq:Lap-vert}
b^m_{1/2}(\alpha) = {{2}\over{\pi}} \int {{ \cos{m\theta} d\theta} \over
        \left(s^2 + p^2\alpha^2 -2\alpha\cos{m\theta}\right)^{1/2}}.
\end{equation}
where $p=1$, $s= 1 + r_0^2/a_{\rm pl}^2$ and $r_0$ is the softening
radius. The interpretation given to this modification is to account in
an approximate way for either a numerically required softening
coefficient \citep{KP93} or (in a slightly different form) the
vertical structure of the disk \citep{Ward88}. In combination, the
improvements of the theory represented in equations \ref{eq:Dstar},
\ref{eq:Dsg} and \ref{eq:Lap-vert} are sufficient to negate the need
for the cutoff function. In general, we will use only this form
(without the cutoff function) in our comparison calculations but will
also make comparisons to the form defined in equation \ref{eq:Lap-Bes}
in order to investigate the changes in the torques they are
responsible for.

Using these analyses, the torque of the disk on the planet will be the
amount of flux injected into or removed from the disk at a given
resonance. For the Lindblad resonances, wave propagation is forbidden
between the ILR and OLR and the flux is zero there. Therefore, we can
calculate the theoretical torque on the planet if we calculate the
angular momentum flux at the resonance, using equation
\ref{eq:LR-flux}. For the corotation resonances, the torque on the
planet may be calculated from equation \ref{eq:CR-flux} as the
difference between the left and right limits as $r$ approaches $r_c$.

The torque, $\delta T$, exerted on any given ring, $\delta r$, in the
disk is accounted for as a torque density, $dT/dr$, as is the torque
actually transmitted to the disk matter. The first form will be a
function of the amplitude and phase of a wave as it propagates through
the disk and in the second will be a function of the dissipation in
the disk as it decays. In the first case, studied by \citet{Ward86},
the torque density will be characterized by an oscillating function of
distance from the planet while in the second \citep{Takeuchi96}, the
torque will instead be a decaying function of distance. Since our
interest lies mainly in the torque of the spiral pattern on the
planet, our torque densities will be characterized by the oscillatory
form. Corotation torques on the other hand are not associated with
waves. Torque densities on the disk for these resonances may be
obtained directly from equation \ref{eq:CR-flux} by differentiation
with respect to $r$.

\subsection{Deriving torques from the simulations}\label{sec:torq-sim}

In order to make comparisons to the torque formulae above, we must
determine the amplitude and phase of each Fourier component of spiral
density structures present in our simulations. As a function of
radius, these amplitudes are 
\begin{equation}\label{eq:patt-amp} 
X_m(r)={{1}\over{\pi}}\int_0^{2\pi}e^{im\theta}X(r,\theta)d\theta,
\end{equation}
where $X$ is the perturbed quantity (in this case surface density) and
$m>0$. With a normalization of $1/2\pi$ for the $m=0$ pattern, the sum
over all $m$ components yields the surface density at radius, $r$, and
azimuth angle, $\theta$. In our discussions below we will normalize
the surface density amplitudes for all patterns to the $\Sigma_0$
(i.e. azimuth averaged) component. The phase of each component is
\begin{equation}\label{eq:patt-phase}
\theta_m = \tan^{-1}\left[{{Im({X_m})}\over{Re(X_m)}}\right].
\end{equation}

Calculation of the torque on the planet is a straightforward
reconstruction of the density due to each pattern in each grid zone
and a sum of the net torque over all grid zones. With a change of
sign, these torques are also those exerted by the planet on the gas in
the disk at each radius, and the summation can be separated into two
partial sums to account for torques from radially inward or outward of
the planet.

\subsection{Connection of the numerical results to physical
systems}\label{sec:connect}

In following discussion, we will show the torque magnitudes and torque
densities exerted by the disk on the planet, either due to accretion
or to gravitational interactions. As a point of reference in
evaluating the figures, recall that the present day Jupiter has a
total orbital angular momentum of $1.9\times10^{50}$~g cm$^2$/s.
Therefore a torque of 10$^{38}$~g cm$^2$/s$^2$ exerted for a time of
10$^3$ years will cause it to migrate inward from 5.2~AU to 5.0~AU.  A
0.3\mj\  planet would move inward to 4.6~AU under the same influence.
When spread over a radial region of 0.25~AU, a torque of this
magnitude yields a torque density of $\delta T/\delta
r\sim7\times10^{24}$~g cm/s$^2$.

\section{The importance of accretion on planet migration}\label{sec:acc-import}

Many workers have shown in linear and nonlinear analyses that a planet
embedded in a disk, especially before a gap has formed, is expected to
have strong dynamical interactions with the disk matter within a few
disk scale heights radially inward and outward of the planet. We found
in \jone\  that in fact the migration was dynamically very sensitive
to the disk mass close to the planets-within one or a few Hill radii
in both radius and azimuth coordinates. One additional interaction
specific only to this region is mass accretion. In the following
sections, we examine the importance of mass and momentum accretion
onto the planet on its trajectory and the formation of a gap using
simulations that allow accretion onto the planet at varying rates. We
then inquire into the physical reasonability of those rates, in order
to better constrain the possible morphology and growth processes of
forming planets. The parameters defining each of the simulations in
these experiments are summarized in table \ref{tab:table-acc}.

\subsection{Accretion}\label{sec:accrete}

We saw in \jone\  that planets with enough mass can open a deep gap in
the disk via gravitational torque interactions and in so doing
drastically slow their migration. Since the disk begins in a somewhat
artificial condition (a 1\mj\  planet should have already either
formed a gap or been accreted by the star) we cannot consider the
rapid motion prior to gap formation typical behavior of the long term
behavior of real systems with such massive planets.

Lower mass planets remain at issue. The migration of a planet less
massive than 0.3\mj\  does not lead to a gap and is fast enough to
move inward on a $<<10^5$~yr time scale, far shorter than the expected
disk lifetime of $>10^6$~yr and the expected formation time scale for
planets. The critical conclusion to note is that low mass planets can
be influenced by gravitational torques from the disk strongly enough
to migrate quickly through it, but cannot influence the disk strongly
enough to form a deep gap and enter the much slower \ttwo\ migration
phase. How then can such objects survive long enough to remain
separate from the central star as the system moves into its main
sequence lifetime?

Accreted matter changes the angular momentum of the planet due both to
the increase of mass and because that matter does not in general have
identical specific angular momentum to the planet when it is accreted.
Additional matter accreting onto the planet continues to bring with it
orbital angular momentum, exerting an effective torque on the planet's
trajectory. If forming planets are to survive for more than a few
thousand years, then at least one of three conditions about this
accreted matter must be true. First, if this angular momentum
contribution is positive (and large enough), then it will provide a
positive torque on the planet, counteracting the negative dynamical
torques acting on it. Second, the planet could accrete all of the
nearby disk matter, thus opening a gap and eliminating the large
gravitational torques driving the migration. Third, mass accretion
onto the planet could proceed faster than \tone\ migration, so that
planets become massive enough, quickly enough to transition to \ttwo\
migration before migrating all the way inward to the stellar surface.
Each of these possibilities can be true only if the disk can supply
matter at a fast enough rate and the planet can accept matter at that
rate. 

\begin{figure}
\psfig{file=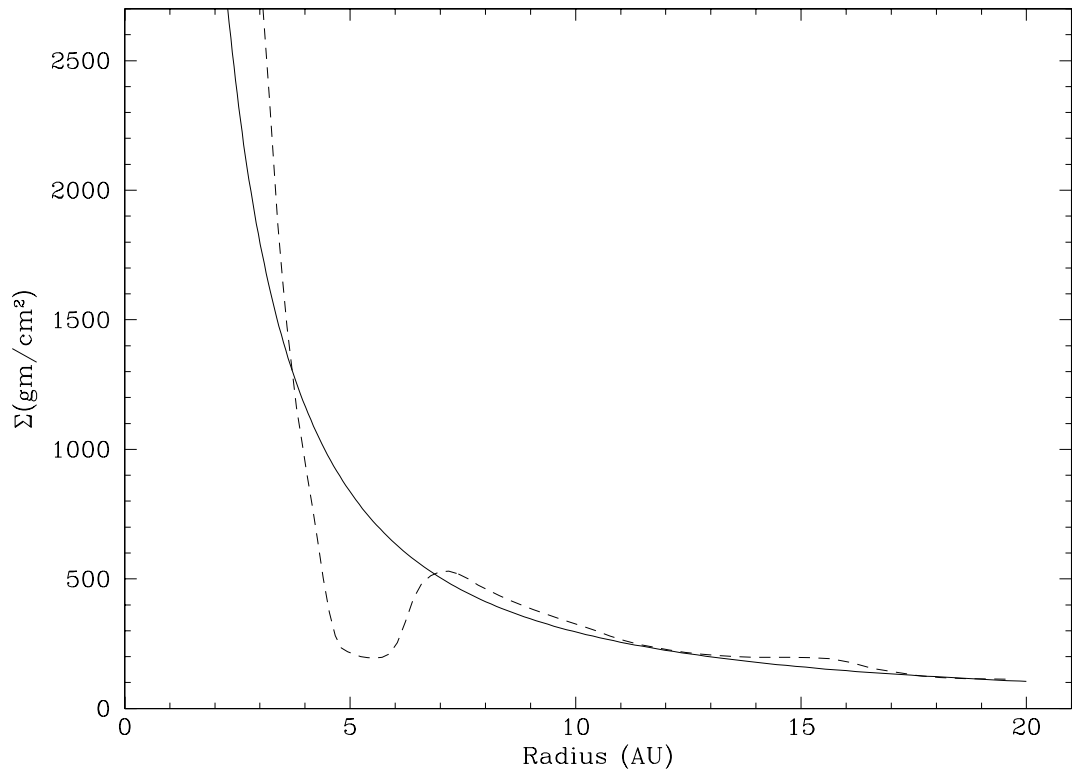,width=15cm,angle=-90}
\caption[Initial surface density for the disk]
{\label{fig:sd-conds}
Initial conditions for surface density of the disk. The dotted
vertical lines denote the inner and outer grid boundaries. The dashed
line defines the azimuth averaged density distribution of the initial
conditions for the {\it gap} series of simulations, defined in the
text.} 
\end{figure}

We have explored the viability of these three conditions with a set of
simulations that include accretion onto the planet, and use our low
mass prototype simulation as a model. In order to test whether the
evolution is merely an artifact of the fact that our initial condition
is an unperturbed disk, we consider two cases. We consider an initial
condition identical to our low mass prototype, and we consider an
initial condition which begins with an already formed gap in the
initial surface density profile around the planet, as shown in figure
\ref{fig:sd-conds}. These two series are designated {\it acc} and {\it
gap} in table \ref{tab:table-acc}, respectively. For each series, we
vary the accretion rate by changing the assumed fraction, $f$, of the
Bondi-Hoyle-Littleton rate (see eq. \ref{eq:BHL-acc}) in different
simulations.

\begin{figure}
\psfig{file=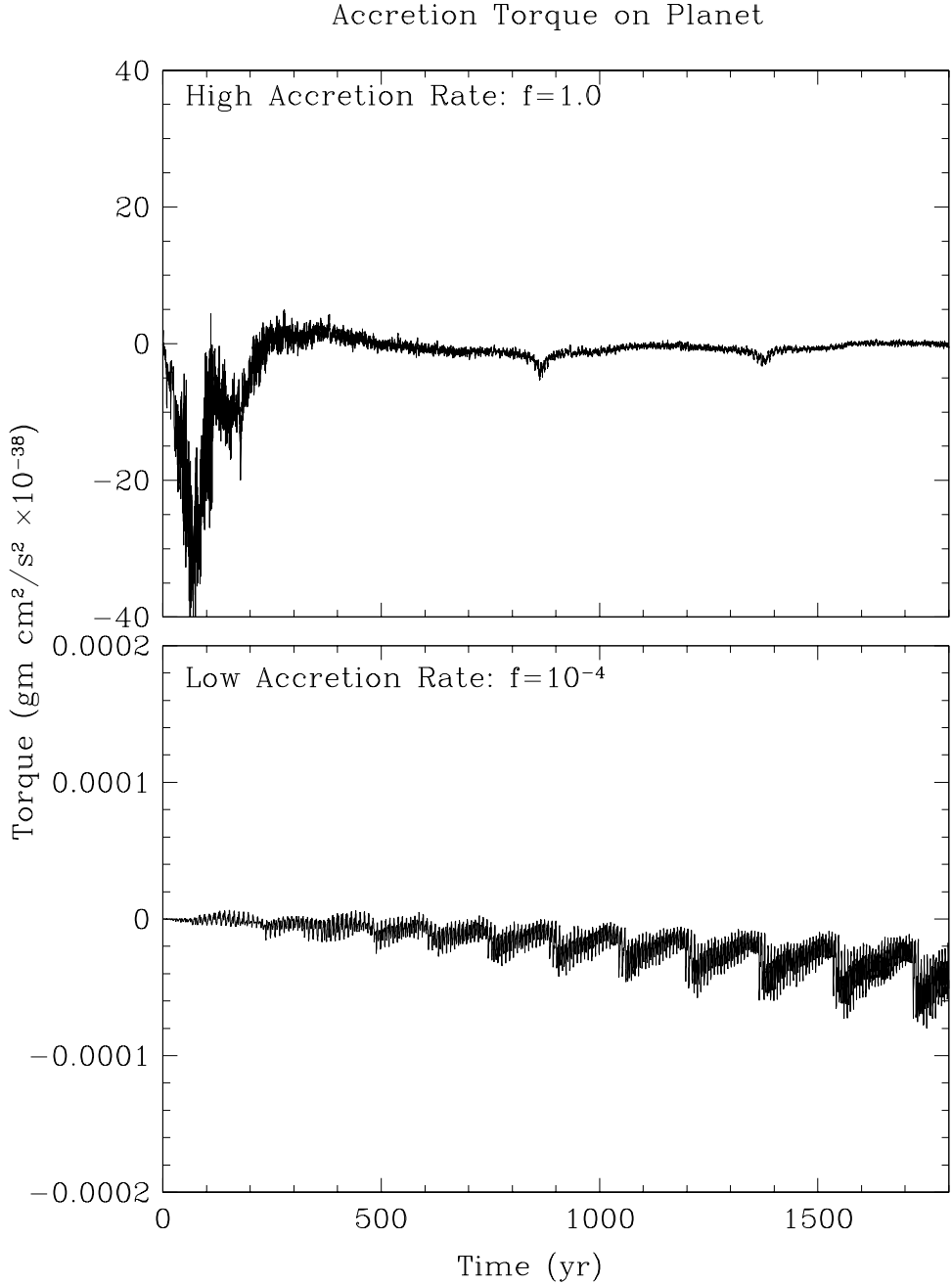,height=7.1in,rheight=6.75in}
\caption[Accretion Torque]
{\label{fig:acc-torq}
Contribution to the dynamical torque due to the accretion of matter
onto the planet. Top: the torque for the no initial gap model, {\it
acc0}, with $f=1.0$. Bottom: the torque for the $f=0.0001$ model, {\it
acc4}. In
each case, although the magnitudes of the torques due to mass
accretion are different, their signs are the same. Mass accretion
provides a net, negative torque on the orbital angular momentum of the
planet. In each case, the torque is smoothed using the average torque
in a moving 0.1~yr window, in order to average out the short term
effects of our finite grid resolution. The periodicity in the {\it
acc4} model is due to the passage of the planet through successive
rings of grid zones as it migrates inward towards the star. Its long
term negative slope is due to the increase in mass of the planet (see
equation \ref{eq:accrete-torq}).}
\end{figure}

\begin{figure}
\psfig{file=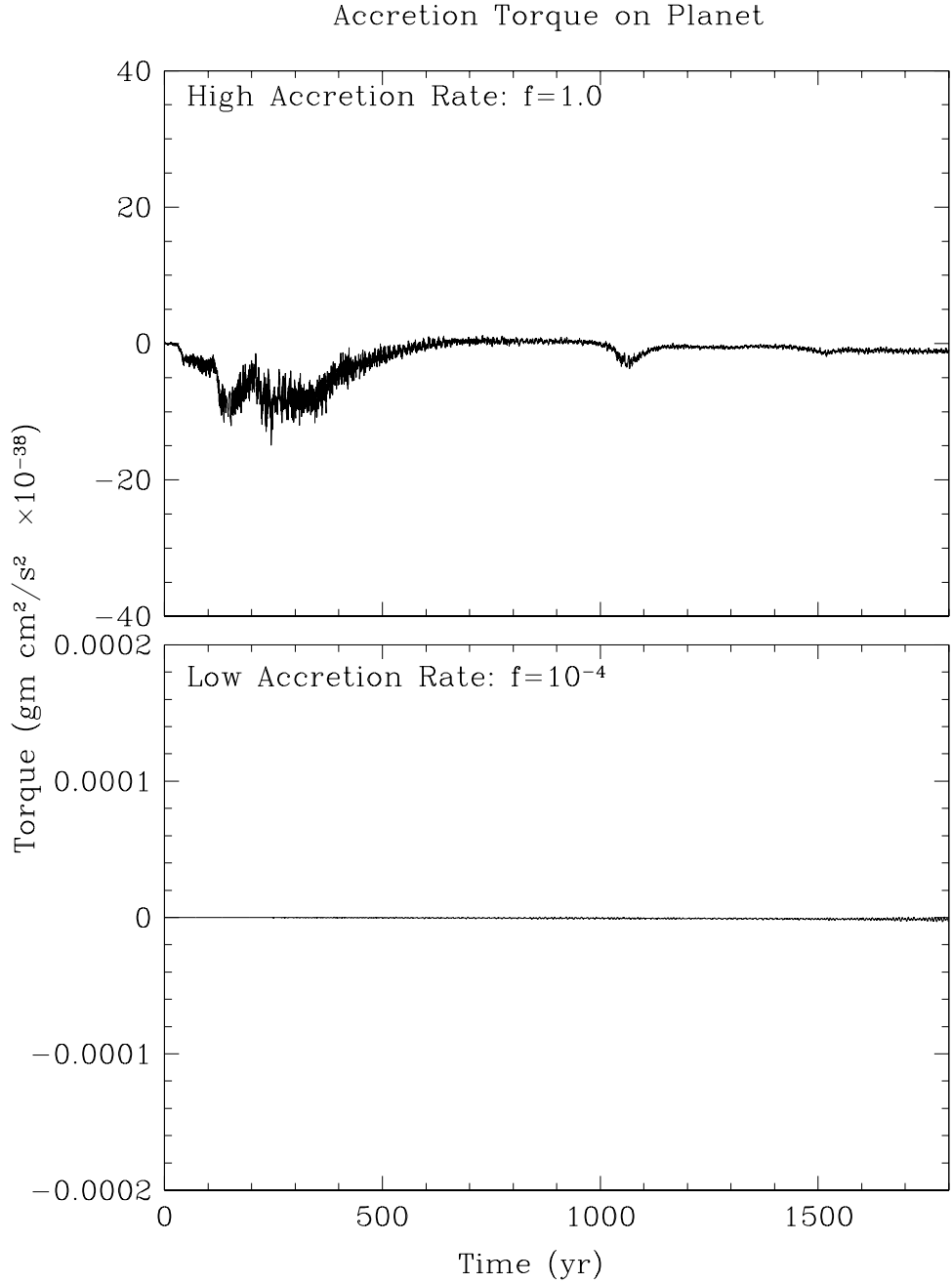,height=7.1in,rheight=6.75in}
\caption[Accretion Torque with an initial gap]
{\label{fig:gap-torq}
Same as figure \ref{fig:acc-torq}, but for the models with an initial
gap. Top: the torque for the initial gap model, {\it gap0}, with
$f=1.0$. Bottom: the torque for the $f=0.0001$ model, {\it gap4}. In
each case, although the magnitudes of the torques due to mass
accretion are different, their signs are the same. Mass accretion
provides a net, negative torque on the orbital angular momentum or the
planet.}
\end{figure}

Is the accretion torque positive and large, satisfying our first
condition? In figures \ref{fig:acc-torq} and \ref{fig:gap-torq}, for
the unperturbed and initial gap simulations respectively, we show the
accretion torque as a function of time for a high accretion rate and
for a low accretion rate simulation. For the two $f=1$ models, the
torque magnitude is initially large compared to the dynamical torques
exerted on the planet by individual spiral patterns (see section
\ref{sec:pattern-ev} below), but within $\lesssim500$~yr drops to near
zero as the planet accretes additional mass. In both examples, the
migration of the planet proceeds inward from 5.2~AU to 4.8~AU and
stops. The low accretion rate models each have accretion torques which
are far smaller than the dynamical torques on the planet. In the {\it
acc4} model, the migration proceeds in near identical fashion to the
low mass prototype simulation, while the {\it gap4} model, the planet
remains at 5.2~AU for the duration of the simulation. The sign of the
torque is always negative. Therefore it acts to increase the planet's
inward migration rate, and we conclude that a planet's migration
cannot be halted by the accretion of orbital angular momentum from the
disk. Our first condition is not confirmed. 

\begin{figure}
\begin{center}
\psfig{file=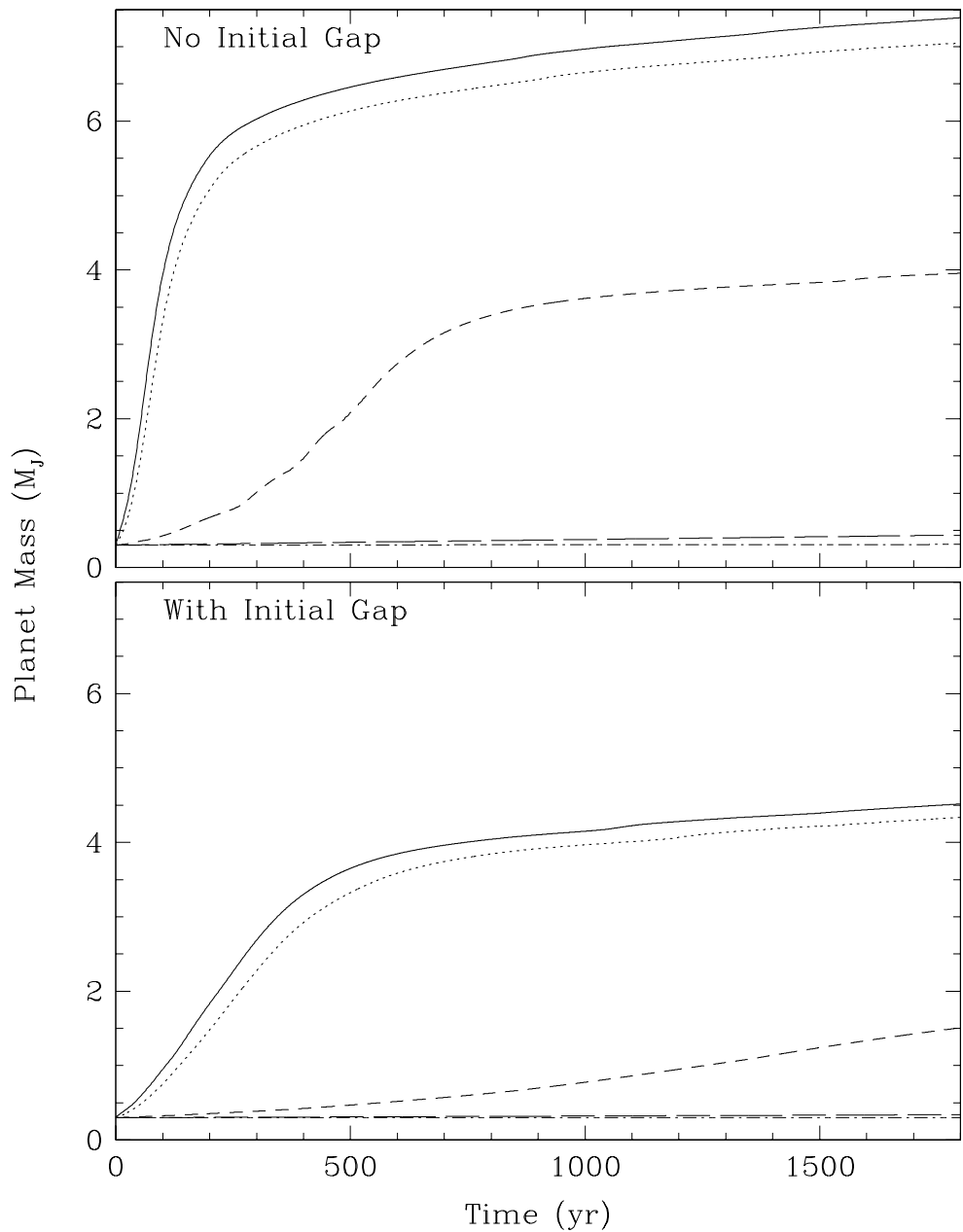,height=7.2in,rheight=6.6in}
\end{center}
\caption[Mass Accretion]
{\label{fig:accrete}
Mass of the planet as a function of time, for different assumed Bondi
accretion rate fraction, $f$. Accretion rates defined by $f=$1.0, 0.1,
0.01, 0.001 and 0.0001 are shown with solid, dotted, short dashed, long
dashed and dash-dotted lines respectively. In the top panel: for each
of the $f>0.001$ curves, the accretion is initially very rapid, then
slows because a gap forms in the disk by virtue of the fact that all
of the mass there is either accreted onto the planet, or driven to
large distances from it via dynamical torques. In the bottom panel:
accretion remains very rapid even in the presence of an initial gap in
the disk. Again, for simulations with $f>0.001$, the planet is able to
accrete $>$1\mj\  within 1-2000~yr.} 
\end{figure}

Can a forming planet can accrete all of the nearby disk, thus forming
a gap and satisfying our second condition? In figure
\ref{fig:accrete}, we show the mass of the forming planet as a
function of time for different fractions, $f$. For fractions,
$f\ge0.01$, the planet accretes between four and eight \mj\  of mass
from the disk in only 1800~yr. The initial accretion rates are as high
as $\sim1$\mj/25~yr in the most extreme cases. The accretion occurs so
quickly that the limiting factor in the accretion is not the BHL rate
given by eq. \ref{eq:BHL-acc}, but rather the fact that the disk
simply cannot supply matter to the planet at the calculated BHL rate
fraction. Instead, the rate is limited by the maximum amount of mass
that we allow to be removed from a given zone per time step. The
region inside the Hill sphere of the planet becomes completely drained
and a gap is quickly generated in the disk as matter continues to pass
into the planet's Hill sphere. As found by \citet{Bryden99} and
\citet{LSA99}, we also find (by virtue of the fact that this region is
the first to be emptied of matter) that matter accretes onto the
planet first from the regions approximately 2-4\rh\  inside and
outside the planet's orbit radius. Matter in the `horseshoe' region,
at the planet's orbit radius but with a phase relative to the planet
greater than $\pm\pi/3$, is able to remain unaccreted for longer
periods, but is also lost within 10$^3$~yr when $f>0.01$. 

The accretion rate decreases only when a wide and deep enough gap has
begun to form in the disk. When such a gap begins to form, a
relatively smaller amount of matter can come into close contact with
the planet, and so become perturbed onto a planet intersecting orbit
and accreted. The planet masses at which this turnoff occurs in the
{\it gap} simulations are a factor of $\sim2$ below those reached in
the {\it acc} series. While it is true that the `final' masses for the
planet at the end of each simulation are smaller by about a factor of
two, the basic conclusion remains. These rates are sufficient to
increase the planet's mass to several Jupiter masses in only a few
hundred years.

For a 0.05\msun\  disk, more than 20\mj\  are available to be accreted
in the region $\pm$~1~AU from the planet, so that the planet would
reach a mass of several tens of Jupiter masses rather than the
$\sim6-8$\mj\  that it reaches before the accretion slows. This means
that the disk does not supply matter to the planet at a rate such that
{\it all} of the nearby disk matter is accreted. Therefore our second
condition is not confirmed. However, planets this massive still do
open a gap and make the transition to \ttwo\  migration via dynamical
processes alone, as we saw in \jone. This means that the disk is
capable of supplying the planet with matter quickly enough to drive
gap formation by dynamical processes, so that our third possibility
may be confirmed if the planet can accept matter at these rates.

Our first two conditions for survivability of Jovian planets (torques
due to accretion and accretion of all available matter) could be
definitely ruled out in this section. While accretion of additional
mass can indirectly drive gap formation by dynamical processes, the
accretion torques have little direct effect on halting or accelerating
the migration, even for the most extreme simulations we performed.
These conclusions are true for both the unperturbed and initial gap
simulations and so are not an artifact of the initial condition. The
third condition (accretion of enough matter to open a gap via
dynamical processes) could not be confirmed or ruled out directly.
However, we can take the information that Jovian mass planets exist
around the sun and other stars, along with the failure of our first
two possibilities, as an indirect confirmation. One significant
ambiguity remains however: how fast can mass accretion onto the planet
actually proceed?

\subsection{The morphology of the circumplanetary environment and its
influence on the planet's accretion rate and final
mass}\label{sec:plan-env}

Deriving limits on the planet's accretion rate is important for both
the survivability of the planet and determination of its final mass.
In this section, we will set a lower limit on the accretion rate from
the condition that it survive long enough to form a gap, and a (weak)
upper limit by examining the morphology of the planet implied by the
accretion. We will conclude that the planet's envelope must undergo
periodic dynamical instabilities during this phase of its evolution.
The efficiency of these instabilities in accelerating the mass
accretion will influence the final mass of the planet. 

We can define a lower limit on the mass accretion rate of the forming
planet from the results \jone. We know that low mass planets migrate
on a time scale of order 0.5--1~AU per thousand years, and can open a
gap when they reach $\sim$0.3\mj. This means that in order to survive
long enough to open a gap and slow their migration rate, planets less
massive than 0.3\mj\ must grow faster than $\sim10^{-4}$\mj/yr. The
limit will be somewhat lower in disks less massive than the \mrat=0.05
ratio implicit in these calculations, but recall that we are
considering a relatively early stage in the planet's evolution when
the disk may still be more massive than the nominal minimum mass solar
nebula. 

How much larger than this lower limit can the accretion rate be? The
efficiency of accretion will depend to some extent on the distribution
and dynamic and thermodynamic conditions of mass very close to the
planet. For the planet masses discussed in our study, any envelope
that existed earlier has grown massive enough that gravitational
contraction and/or ejection of some of the matter has begun
\citep{BP86,Wuch91}. Eventually the forming planet will begin to
develop a central core$+$disk structure. If we assume a disk structure
has already formed, we can derive a self-consistency check of this
assumption from the gravitational potential energy that must be
radiated by the gas before it is accreted onto the planet and from a
measure of the disk's thickness.

For a given accretion rate, and independent of the specific form of
dissipation in the disk, the central temperature for a steady state,
internally heated accretion disk is given by 
\begin{equation}\label{eq:t-mdot}
T(r) = \left({{9GM_{\rm pl}\dot M_{\rm pl}\tau}
                   \over{32\pi\sigma}}\right)^{1/4}  r^{-3/4}
\end{equation}
where $\tau$ is the optical depth to the midplane and $\sigma$ is the
Stefan-Boltzmann constant \citep[][ch. 5]{FKR}. This formula is
general in the sense that it assumes only that the disk is optically
thick and that the accretion rate is constant. 

A dimensionless measure of the scale height is $H/r=c_s/(r\Omega)$. If
we assume that $H/r=1$ (a serious violation of the thin disk
criterion), then we can convert the scale height into a temperature
\begin{equation}\label{eq:hr-temp} 
T(r) =  \left({{\mu m_{\rm pr} G M_{\it pl}}\over {\gamma k_B}}\right)r^{-1}
\end{equation}
where $m_{\rm pr}$ is the mass of the proton, $k_B$ is Boltzmann's
constant, $\mu$ is the average molecular weight of the gas and
$\gamma$ is the ratio of specific heats. In order for the
circumplanetary environment to be `disky', gas temperatures must be
well below the limits defined by eq. \ref{eq:hr-temp}.

For what accretion rates are such temperatures obtained in the
circumplanetary environment? In order to answer this question we
require the values of the average molecular weight and the ratio of
specific heats which, for conditions appropriate for circumplanetary
disks, will be $\mu\approx2.3$ and $\gamma\approx 1.4$. We also
require  a value for the optical depth, $\tau(=\Sigma\kappa$). We can
make a very conservative estimate of the surface density, $\Sigma$, if
we assume that the circumplanetary disk mass at given time is the mass
of the present day Jovian moon system, which we will assume at the
time of formation to consist of both gas and solids in solar nebula
proportions. Then the surface density is $\Sigma\sim 10-20$~g/cm$^2$
for a disk extending outwards to the Hill radius. For a solar
composition, the Rosseland opacity of this matter at typical nebular
temperatures will be $\kappa\sim2-4$cm$^2$/g \citep{Pol94}, so that
the optical depth is $\tau=\Sigma\kappa\sim10-100$.

\begin{figure}
\psfig{file=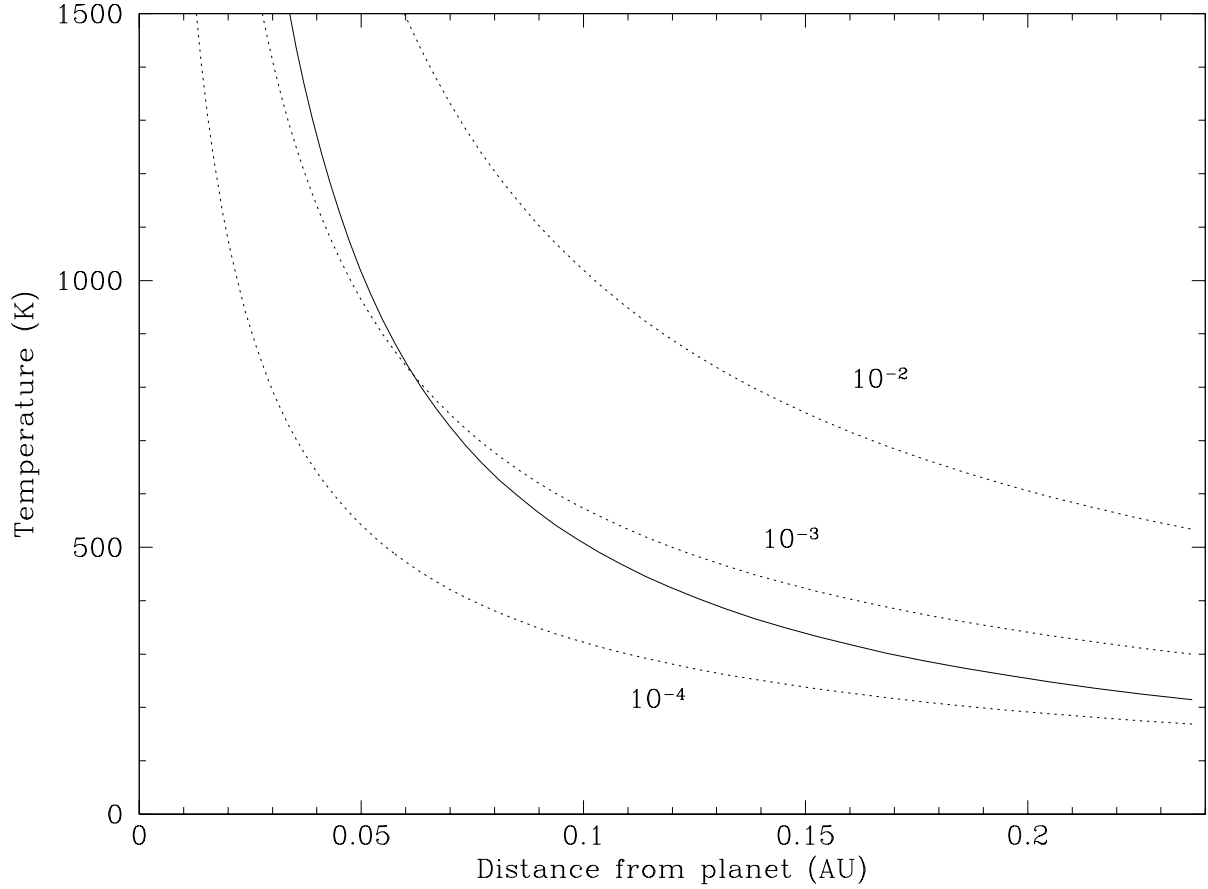,height=5in,angle=-90}
\caption[Disk Temperatures]
{\label{fig:accdisk-t}
Temperature as a function of radius for an assumed steadily accreting
circumplanetary accretion disk with optical depth $\tau=100$ and
various accretion rates, $\dot M_{\rm pl}$ (dotted curves). With this
assumed optical depth, the curves correspond to accretion rates of
10$^{-2}$, 10$^{-3}$ and 10$^{-4}$ \mj/yr from top to bottom. The
temperature defined by the condition that $H/r=1.0$ is also plotted
with a solid curve. Only accretion rates of $\dot M_{\rm
pl}\lesssim10^{-4}$\mj/yr produce temperatures which do not severely
violate the condition that the disk be thin. }

\end{figure}

For an optical depth of $\tau=100$, equations \ref{eq:t-mdot} and
\ref{eq:hr-temp} yield curves as plotted in figure \ref{fig:accdisk-t}
for a 0.3\mj\  planet. The condition that the disk have $H/R=1$ is
defined by a temperature of $T\approx 200$~K at a distance of one Hill
radius (0.24~AU) from the planet and reaches the destruction
temperature of silicate dust ($T\approx 1200$K) at a distance of
0.04~AU from the planet ($\sim 40-50$\rplan\ if \rplan=2\rj). For all
accretion rates $\dot M_{\rm pl}>10^{-4}$\mj/yr, the condition that
the disk be thin is severely violated in all but the innermost parts
of the circumplanetary disk. Lower mass planets will have curves with
similar characteristics, but with lower temperature scales, through
the \mplan$^{1/4}$ and \mplan$^1$ proportionalities in the temperature
laws, eq. \ref{eq:t-mdot} and \ref{eq:hr-temp}, respectively. 

Coupling this result to the lower limit on the accretion rate, we
conclude that survival of the planet is inconsistent with the
assumption of a circumplanetary disk at this stage of the planet's
evolution. For most of the time that it is growing, the planet must
accrete matter faster than $\dot M_{\rm pl}>10^{-4}$\mj/yr. This
conclusion places a strong constraint on the planet's morphology. It
must be characterized by an envelope or thick disk structure.

We can use this constraint to derive a weak upper limit to the
accretion rate. The most liberal upper limit is that at which
radiation pressure suppresses accretion of additional material.
Assuming that the opacity source is ionized hydrogen atoms, then this
rate is the Eddington accretion rate, which for an accreting surface
at a radius $\sim1-2$\rj, is $\sim10^{-1}$\mj/yr. However, the
Eddington rate is probably not relevant for accreting Jovian planets
in circumstellar disks, because the main source of opacity is not
ionized hydrogen, but rather dust or exited states of complex
molecules. If we instead assume that the main opacity source is dust
and that the gas and dust remain well mixed, then the appropriate mass
opacity is the $\kappa\sim2-4$g/cm$^2$ value noted above rather than
the $\kappa\sim0.4$g/cm$^2$ appropriate for Eddington accretion, and
an analogue of the Eddington rate would be a factor ten smaller.

A much more restrictive limit than radiation pressure is the
requirement to overcome the gas pressure of the planetary envelope
composed of the recently accreted matter. This rate is available from
an analysis of the Kelvin Helmholtz contraction of the envelope.
\citet{Bryden00} define the maximum accretion rate allowed by
Kelvin-Helmholtz contraction of the envelope as 
\begin{equation}\label{eq:Bry-KH}
\dot M_{\rm KH} =  M_{\rm KH0} \left({{ M_{\rm pl}}\over{M_\oplus}}\right)^4
\end{equation}
where $M_{\rm KH0}=7\times10^{-11}$\me/yr. For a planet with mass
0.3\mj, we derive an accretion rate of $\sim6\times10^{-3}$\me/yr, or
$\sim2\times10^{-5}$\mj/yr. For a planet of half as massive, the rate
is $\sim10^{-6}$\mj/yr. Although \citet{Bryden00} discuss a number of
factors that can affect this accretion rate, none are likely to change
it by the factor 100 or more required to grow faster than the
migration. Therefore we have an {\it upper} limit for mass accretion
which is substantially lower than our {\it lower} limit.

Even ignoring the Kelvin-Helmholtz contraction, the upper limits from
radiation pressure alone are already smaller than the accretion rates
obtained for our models with accretion factors $f=1.0$ and $0.1$,
eliminating them from the parameter space of systems with physically
self-consistent evolution. Elimination of these models is significant
because we can conclude that the accretion rate onto forming Jovian
planets is driven primarily by conditions near the planet, and not by
conditions in the disk. The gas accretion phase of planet formation,
in which the planet accretes the largest fraction of its final mass,
is therefore relatively insensitive to the disk mass except insofar as
it sets the migration rate through the disk. This conclusion has
rather far reaching consequences because it is in essence a statement
that Jovian planet formation and the initial mass function for Jovian
planets are sensitive to the small scale physics in the forming
body--physics such as the planetary spin and opacity (metallicity)
which are common to all such systems--and much less sensitive to the
global thermodynamic or physical properties of the star$+$disk system.

The rates derived from radiation and gas pressure both define upper
limits to the accretion rate onto the planet. We characterize these
limits as `weak' because they are dependent on assumptions that are
probably not met in practice. One assumption that is particularly
dubious is the neglect of the planet's spin and spin accretion rate in
our estimates. We find that many times Jupiter's present spin angular
momentum is accreted during the 1800~yr of evolution followed in our
simulations. Undoubtedly this value is inaccurate--due both to our low
resolution and to our neglect of angular momentum exchange between the
planet and disk, but it serves to draw attention to one point. If
rotation plays an important dynamical role in the growth, then
dynamical instabilities may develop in which the envelope becomes
unstable to spiral structure growth and rapid mass transport.

In one dimensional calculations, \citet{KBP91} showed that when
exchange between the envelope and circumstellar disk is allowed, only
$\sim2-3$\% of the angular momentum remained in the planetary system,
with the rest immediately returned to the solar nebula. Since their
calculations were one dimensional, they could make no statements about
the nature or mechanism for this transport however. Rotational
instabilities are known to develop in morphologically similar entities
such as rotating polytropes \citep{PDD,TIPD}. We speculate that the
mass accretion rate in forming Jovian planets could be enhanced if
large scale dynamical instabilities in the planet's envelope could
develop, due to this spin accretion and the processes that return it
to the solar nebula. Clearly, much more work is required before a more
firm conclusion can be drawn on this point.

\section{The dynamical interaction of the planet and disk}\label{sec:interact}

In \jone\ and in section \ref{sec:acc-import} we studied the planet
and the disk somewhat qualitatively in terms of their actions on each
other and the consequences for the evolution. We found that some of
the major qualitative predictions of analytic theories were only
partially recovered in our simulations. Here, we will attempt to make
the analysis more quantitative in terms of the applicability of
analytic formalisms to the system, and in what limits they break down.
Because of the large and important differences in outcome, depending
both on grid resolution and on whether disk self gravity was included
or not, we will do a side by side analysis of four models from \jone:
the low mass (with our `standard' resolution of 128$\times224$) and
high resolution prototype models with and without disk self gravity.
The standard resolution (in the low mass prototype) vs. high
resolution comparisons will provide insight into the role of the mass
distribution in the torque calculations, while the self gravitating
vs. non self gravitating models provide insight into the physical
model itself. In a sense, these comparisons will also touch on the
differences between gap forming and non gap forming systems, since the
two models with self gravity form gaps, while those without it do not.

In sections \ref{sec:pattern-ev} and \ref{sec:torq-comp}, we show the
torques from our simulations, first as a function of radius, then as a
function of Fourier pattern number, $m$, and compare them to the
analytic predictions. Then in section \ref{sec:examination} we examine
the approximations made in deriving the analytic torque formulae and
the approximations made in our numerical realization of the system.
While very important for mathematical models of migration, this
discussion will be rather detailed and of lesser interest to some of
our audience. These readers may safely skip forward to section
\ref{sec:torq-sens}, where we discuss the consequences of failures of
various mathematical assumptions on the gravitational torques when
they break down.

\subsection{The torques exerted by the planet and disk on each 
other}\label{sec:pattern-ev}

In figure \ref{fig:torqrad-tot}, we show plots of the gravitational
torque density of the disk on the planet as a function of radius for
the low mass and high resolution prototype simulations. The torques
densities shown are those at the same time (300~yr), as in the top
panels of figures 4 and 6 of \jone, for which spiral patterns have
fully developed and have had time to propagate through the entire
disk, but before a deep gap has formed. These torques will be
representative of those expected from \tone\  migration assumptions.

In each case, the torque density is very large near the planet
(radially) and decays as a function of distance inwards and outwards
from the planet, as the spiral patterns themselves decay.
Qualitatively, the morphology present is consistent with the
theoretical model that low order $m$ patterns contribute little to the
total torque, and that higher order $m$ patterns, whose resonances
fall closer to the planet, are excited and provide most of the net
torque contribution. The sign of the torque oscillates so that at some
radii it acts to increase the disk's angular momentum at the expense
of the planet, while at other very nearby locations it acts in the
opposite sense. The oscillations are stable in time relative to the
planet's position. The torque curves shown with the dotted line
(omitting the contribution from inside the Hill sphere) shows that
while the matter inside the Hill sphere makes some contribution to the
total, it is not by itself the determining contribution to the torque
from this radial region. 

The contribution to the total due to the $m=1, 2$ and 10 spiral
patterns are shown in figures \ref{fig:torqradm-lo} and
\ref{fig:torqradm-hires}, for the same four models. In each case, a
torque oscillating in sign originates at an LR and extends with
decreasing amplitude in the direction away from the planet. Consistent
with our resolution dependent numerical dissipation, the lowest order
(longest wavelength) patterns propagate the furthest distances from
the planet and produce torques over nearly the entire radial extent of
the disk, while higher order patterns contribute to the torques only
very close to the planet. The low order $m$ patterns each display a
large torque near the planet as well, clearly distinct from the
oscillating torque pattern further away and presumably due to the
corotation interaction.

The $m=1$ pattern represents a special case pattern--it has no inner
Lindblad resonance. Therefore we expect that no spiral structures
should be generated there and no torque on the planet from interior to
its orbit due to this pattern should exist. Indeed this is the
case--the torques from the $m=1$ pattern from well inside the planets
orbit are near zero, while a decaying wave structure in the exterior
disk is visible. The $m\ge 2$ patterns do have both inner and outer
Lindblad resonances and for the $m=2$ pattern, the torque oscillates
in sign and decays as a function of distance from the planet, while
for the higher order symmetry $m=10$ pattern only the first wave
maximum can be observed.

In each case, the torques and their oscillations are larger in the non
self gravitating disks than in the self gravitating versions. This is
most likely a consequence of the difference in the migration rates and
their effect on the forming gap. Although we have attempted to examine
a point in time before substantial evolution has occurred, gap
formation has begun to reduce the surface density near the planet,
unfortunately causing the torques to be decreased by varying amounts
in each simulation. While slightly visually disturbing, the
differences will have few if any consequences for our comparisons
below. 

Each pattern also produces a large torque contribution from locations
radially very close to the planet, near the corotation resonance
locations and where waves generated from Lindblad resonances are
forbidden. For all simulations except the high resolution non self
gravitating version, the contribution is positive in sign (increasing
the planet's angular momentum) both inside and outside the planet's
orbit, but negative in sign (decreasing its angular momentum),
slightly further inward. These torques are resolved on our grid in the
sense that they are distributed over many radial rings of grid zones,
however they differ greatly in character between simulations at
different resolution. Therefore, we believe that their true character
is not fully resolved by our simulations. 

\subsection{Comparison to linear analyses}\label{sec:torq-comp}

\begin{figure}
\psfig{file=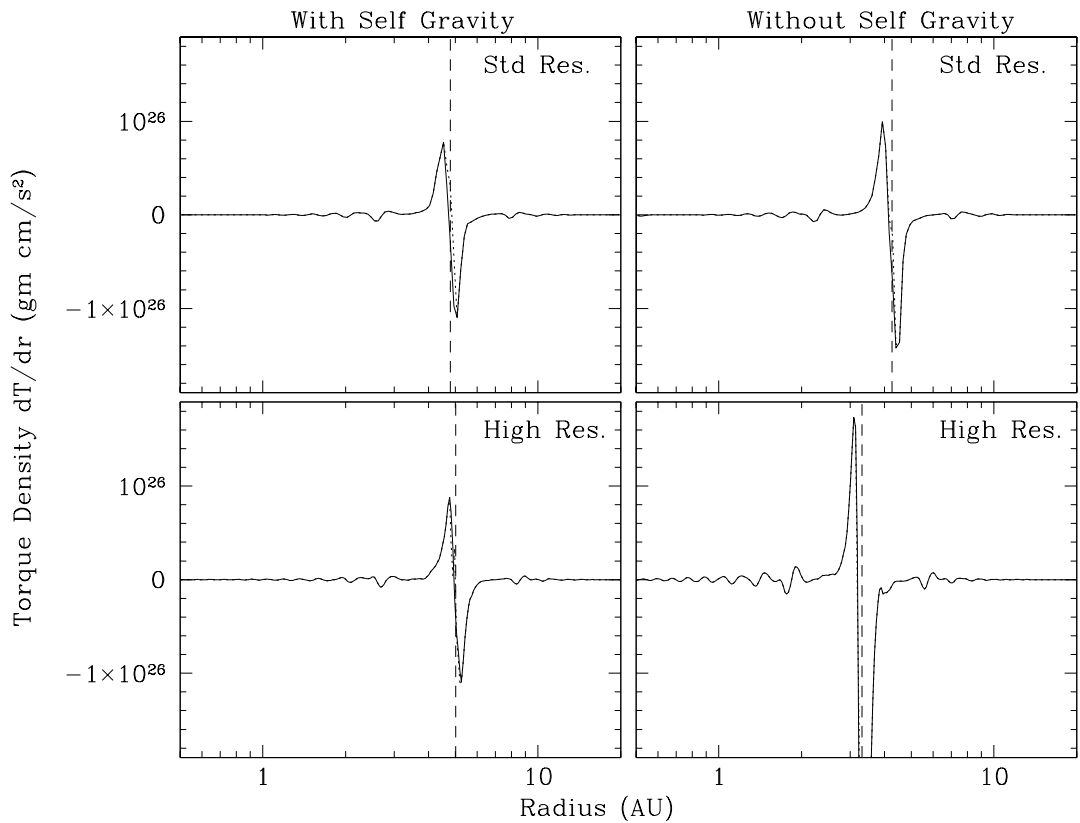,height=4.5in,rheight=4.5in,angle=-90}
\caption[Torque on disk]
{\label{fig:torqrad-tot}
Torque density of the disk on the planet as a function of distance
from the star, for the self gravitating (left) and non self
gravitating disk versions of the low mass (top panels) and the high
resolution (bottom panels) prototype simulations, as indicated. Each
panel shows the total torque (solid line) and the total torque
omitting the contribution from within the Hill sphere (i.e. within
1.0\rh) of the planet (dotted line). The dotted and solid lines
are nearly coincident in the plot, showing that the Hill sphere itself
contributes relatively little to the planet's motion. The orbit radius
of the planet is shown with a vertical dashed line in each panel. The
time at which the torque is evaluated is $t=300$~yr after the initial
time for each simulation. } 
\end{figure}

We have seen above that the torques on the planet can include some
contribution from locations far from resonance locations, and torques
from near both the corotation and Lindblad resonances. Cursory
examination of figures \ref{fig:torqrad-tot}--\ref{fig:torqradm-hires}
suggest that some characteristics predicted by theory for those
resonances may not be identically reproduced in the simulations (e.g.
we would expect from equation \ref{eq:CR-flux} that the torque
near CR would have the same sign over the entire radial range where it
is exerted, but this is rarely the case). Nevertheless, we will
proceed by identifying the torques from the region around each
resonance location, with the interaction predicted theoretically for
that resonance. Separating these torques from each other in our
simulations however, proves to be a challenging problem because for
higher order $m$ patterns, the Lindblad resonances are found
progressively closer to the planet.

For the lowest $m$ patterns, a clear separation exists between the CR
and LR torques because their positions are well separated from each
other. Therefore, for the purposes of our study, we shall define the
LR torques as those exerted in the region between $r=\infty$ ($r=0$)
and the point half the distance between the Lindblad resonance and the
corotation resonance for the OLR and ILR torques respectively. This
allows us to include the portion of the torque which may be produced
at some distance from the resonance itself, but nevertheless is clearly
associated with it. We arbitrarily assign the difference between the
total torque and the sum of the two LR contributions to the corotation
resonance. For higher $m$ patterns (above $m=15$), where all three
resonance locations are very close together and it becomes impossible
to distinguish between the torques from the LRs from the corotation
torque, we shall assign the entire torque to the LRs.

\subsubsection{Torques from near the Lindblad Resonances}\label{sec:torq-LR}

The contributions to the torque from each Fourier component and due to
the Lindblad resonances are shown\footnote{Here and throughout, we
shall plot the torques as continuous functions of the (integer valued)
Fourier pattern number $m$, rather than plotting their values as
points, in order to enhance readability of the plots.} in figure
\ref{fig:mtorq-LR} and \ref{fig:mtorq-LR-nosg} for the models with and
without disk self gravity. We also show the torques broken down to
show the contribution from inside and outside the planet's orbit.
Differences exist, but in general, many qualitative features obtained
for the different runs are similar to each other. In each simulation,
the dominant contribution to the torques comes from the spiral
patterns with $10<m<20$ and as $m$ increases, the torque magnitude
contributions decrease monotonically to zero. Below $m=20$, the total
LR torque may change by a factor of two or more between one pattern
and its neighbors of higher or lower symmetry. For most patterns, the
total Lindblad torque is negative, but this is not universally the
case. In each of the simulations for example, positive total torques
are present for the patterns near $m\sim5$.

While the inner (positive) torques are quite similar to each other in
all four simulations, the outer (negative) torques are larger in the
non self gravitating disks than in the self gravitating versions,
especially for the higher $m$ patterns. This is consistent with the
result that planet migration proceeds more rapidly in absence of disk
self gravity. In the high resolution versions, there is more variation
between the torques of individual patterns, probably because of the
improved ability to distinguish them from the CR torques.

Also shown in figures \ref{fig:mtorq-LR} and \ref{fig:mtorq-LR-nosg}
are the torques predicted from equation \ref{eq:LR-flux}. To derive
these torques, we use the azimuth averaged surface density and
rotation curves of each simulation as inputs and include the
generalized resonant denominator of equations \ref{eq:Dstar} or
\ref{eq:Dsg} both to determine the resonance positions and in the
torque formula itself. We also use the generalized Laplace
coefficients of equation \ref{eq:Lap-vert}, with the same softening as
is used in the simulations. Some qualitative features are reproduced
well by the theoretical calculations. In each case, the torques are
small for both small and large $m$, with maxima in both the net and
inner and outer torque contributions near $m=10$. The positive net
torques near $m\sim5$ are also reproduced.

There are also several very serious differences between the torques from
theory and from simulation. The most significant is that in each case,
the torques from theory are systematically larger by as much as a
factor of six, than those from the simulations. Secondly, the torques in
the non-self gravitating versions are larger than those in the
corresponding self gravitating version. In three of four simulations,
the torque distributions from theory are smoother as functions of $m$
in the sense that variations between nearby $m$ patterns are small.
The exception is the high resolution self gravitating model, which
produces a large, negative net torque at $m=8$ and a large increase in
both the inner and outer torques near $m=10$, gradually decreasing as
$m$ increases.

\subsubsection{Torques from near the Corotation Resonances}\label{sec:torq-CR}

Figure \ref{fig:mtorq-CR} shows the total torques on the planet from
the corotation resonances with $m\le15$. In each simulation, the most
striking feature is that many patterns produce torques that are as
large or larger than the net contribution from the LRs. The
contributions for the patterns with $m\lesssim3$ are particularly
interesting because not only are they large, they are also positive,
thus acting to slow the inward migration of the planet. For patterns
with $m>3$, the torques are negative and again significant in
magnitude compared to the torques from the LRs. Only for patterns with
$m>10$, do the torques decrease to near zero and only when disk self
gravity is present. The net torque from the CRs with $m\le15$ is about
a third the value of the torque from the LRs.

After the simple observation that many of the CR torque components are
large, we are confronted with the uncomfortable situation that large
differences that exist between the results of each of the simulations,
including differences between the simulations which differ only in
resolution. The most significant differences are found in the
magnitude of the torque from the $m=1$ pattern. It is larger when self
gravity is not included and it increases by nearly a factor of two
when we increase from our standard to high resolution.

In both the standard and high resolution versions without self
gravity, the torque from $m=1$ was so much larger that it could not be
displayed on the plot while also displaying features from $m>1$. The
same was true in figure \ref{fig:torqradm-hires} for the radial
distribution of the torque. More important to note than its exact
value is that the contribution is more than three times the largest
one sided LR torque (i.e. ILR or OLR). It also has no counterbalancing
torque of opposite sign as the LRs do, making its contribution very
important in a determination of the migration rate. Further, for the
high resolution simulations (but not the standard resolution models),
the torque is large and positive when disk self gravity is included,
but large and negative when it is suppressed.

Higher order $m$ patterns also display differences. While the standard
resolution simulations have near zero CR torques, both of the high
resolution counterparts have large negative torques. In this case,
both high resolution simulations produce negative torques for
$5<m<10$, so that the inward migration is more rapid. Above $m=10$,
only the non-self gravitating simulation produces negative torques.

Because of the large and qualitative differences between the
simulations, we are forced to conclude that we have not adequately
resolved the effect that the CR torques will have. We have therefore
not attempted to make a direct comparison of the CR torques with their
analytic predictions, as we have for the LRs. While we cannot make
reasonable comparisons with theory, we may still draw important
physical conclusions from the large size of the CR torques. Namely,
that they may have a much larger effect on migration than previously
realized. In order to constrain this possibility further, extremely
high resolution global simulations of disks during the \tone\
migration stage must be performed. 

\subsection{Examination of the Assumptions and Approximations Made in
Linear Analyses}\label{sec:examination}

What is the origin of the serious differences between theory and
simulation, and between one simulation and another? In this section,
we will examine the effects of the variation of the rotation curve on
the torque and small shifts in the resonance locations may have on the
calculated torques, and in the following section, their significance
for the torques. Such shifts may be global in nature, for example a
shift in the initial state due to self gravity or pressure forces.
Variation in the rotation curve may also be more local in nature, for
example the changes in the rotation curve due to steep pressure
gradients at the edges of the forming gap. Except for the discussion
of linearity in section \ref{sec:linearity}, in the following sections
we will concentrate specifically on our high resolution prototype
simulation (with and without self gravity) so that these small shifts
may be determined more precisely.

\subsubsection{Validity of the resonance
approximation}\label{sec:res-approx}

The wavelike behavior of the lowest order patterns is a consequence of
the fact that at some radii, the spiral patterns have a different
phase relative to the planet than at other radii. The torque exerted
on the planet by the disk matter at that radius is therefore positive
or negative depending on this phase. The important points to note are
that the torque is exerted at locations significantly different from
the Lindblad resonance and that the sign of this local torque can be
opposite that of the prevailing flow of angular momentum. 

How much do the torques exerted far from the resonances contribute to
the total? This question is important because one simplifying
assumption made in deriving equations \ref{eq:LR-flux} and
\ref{eq:CR-flux} is that the disk and planet interact only {\it at}
the resonances, rather than at some distance away from it. Errors in
the comparison will enter if torques far from resonances contribute a
significant fraction of the total.

Based on inspection of figures \ref{fig:torqradm-lo} and
\ref{fig:torqradm-hires}, the simulation most strongly affected by
torques far from the resonance will be the non self gravitating, high
resolution version because its torques have the largest amplitude
oscillations everywhere. For this model, figure \ref{fig:cum-torq}
shows show the cumulative sum of the torque as a function of distance
from the planet (both inward and outward) originating from the
Lindblad resonances, Torques from corotation interactions are
suppressed. While some errors are made due to incorrect separation of
the CR and LR contributions, in every case the cumulative torque
magnitude increases to its maximum value slightly outside or inside
the exact resonance position for the OLR or ILR respectively. At
greater distances, it undergoes decreasing amplitude oscillations
around what becomes its final value at large distances from the
planet. In every case the oscillations are smaller in amplitude than
in the first 1/2 cycle of the wave. Therefore, the largest fraction of
the net torque from a given resonance is indeed derived from near the
resonance position itself.

The maximum cumulative torque is never obtained at the resonance, but
rather slightly further away from the planet, after which the
cumulative sum falls to a value as much as 30--50\% below its initial
maximum. This is to be expected since the torque density waveform can
be approximated as an Airy function \citep{Ward86} whose first maximum
is found about 1/4 cycle more distant from the planet than the
resonance. The correspondence is also strengthened by the fact that
the integral (i.e cumulative sum) of the Airy function also drops
nearly 40\% from its initial maximum during the next 1/2 cycle of the
wave. 

A useful visual diagnostic for comparison between patterns is the
position of the first maximum of the cumulative torque relative to the
resonance position. Using this measure, there is variation between one
pattern and another. For the $m=2$ pattern for example, the ILR
position is at approximately the half maximum, while the OLR position,
even accounting for the positive offset from misattributing a portion
of the CR torque to the LR, is found nearly at the `foot' of the wave.
Variation similar to that shown are present in all other patterns as
well.

We conclude that the resonance approximation of GT79 is partially
supported by our simulations for at least the Lindblad resonances,
since most of the torque between the planet and the disk originates
from the first cycle of the wave. Our method of separating the CR and
LR torques already constrains the CR torques to a narrow region, so we
can make no statements about these contributions. The specific
approximation that the conditions at the {\it exact} resonance
position can be used to determine the torque may not be as well
justified, because the maximum cumulative sum may be found at varying
distances from the exact resonance. 

\subsubsection{Linearity}\label{sec:linearity}

In \jone we found that the migration rates were nearly flat as a
function of planet mass, varying by less than a factor of two over a
factor 20 change in mass. In contrast, theory predicts that the rates
will scale linearly with planet mass. The theory is based on the
assumption that the perturbations are small, so that the models are in
the linear regime. In this limit, we expect the perturbations to scale
with the mass of the perturber, and through them also the migration
rate. If instead perturbations are large, they will saturate--larger
perturbers will not produce larger perturbations. 

Are our migration rates flat because the perturbation amplitudes are
saturated? In figure \ref{fig:maxpatamp}, we show the maximum pattern
amplitudes as a function of mass (i.e. for the {\it mas} series of
simulations--see \jone), obtained in the regions defined for each of
the three resonances, each obtained at a time 300~yr after the
beginning of the run. We again show the $m=1,2$ and 10 patterns as
typical representatives of the behavior of each of the other low and
high order symmetry patterns.

The $m=1$ and $m=2$ pattern amplitudes may reach as high as 40--50\%.
Even with these very large amplitudes, a close correlation between the
relative amplitudes of each of the three resonances with each other is
maintained. Although we have not attempted to fit linear functions
to the data, a clear linear increase is present over most of the mass
range. For example, the amplitude near the $m=1$ OLR is $\sim$2\% at
0.1\mj\  and increases by a factor ten to $\sim20$\% as the planet mass
increase by a factor 10 to 1.0\mj, a perfect linear dependence. The
direct proportionality continues even to the 2\mj\  simulation with an
amplitude 20 times that found at 0.1\mj. The maximum amplitude near
corotation displays similar characteristics. For the $m=2$ pattern,
the maximum amplitudes near the ILR and CR increase in direct
proportion to the planet mass, but the amplitude near the OLR
saturates at $\sim20-25$\% for simulations with planets above 1\mj.

On the other hand, higher order patterns typified by $m=10$ produce
much different behavior. The amplitudes are not as large as for the
$m=1,2$ patterns, and they no longer increase in linear fashion over
the whole range. Instead, above $\sim0.75$\mj, the close correlation
between the amplitudes for each of the three resonances is lost and
the growth with planet mass appears to have saturated. Even below
0.75\mj\  the increase is no longer directly proportional to planet
mass. For example, a 2\% perturbation increasing only a factor four
with a factor 7.5 increase in planet mass. The mass at which
saturation occurs is also similar to that for which the migration rate
shown in figure 9 of \jone\  undergoes its only real change in its
behavior, from a slowly growing rate, to a completely flat function of
planet mass. We conclude that the reason for this change is the
saturation of the pattern amplitudes for the patterns most strongly
affecting the migration.  

At the low end of the plot of mass vs. migration rate from \jone
(figure 9), an extrapolation of the migration rates to zero planet
mass yields a non-zero rate. Such a phenomenon would appear to be
either inconsistent with the view that the torques are dominated by
Lindblad resonance interactions in the linear regime, or with the view
that gravitational torques are responsible for the migration (since a
zero mass planet would not generate such torques). Neither alternative
can be fully supported. Instead, we favor the view that the
simulations of the lowest mass planets (below $0.3-0.5$\mj) are in the
linear regime, but that the true linear increase in the migration rate
is not observed because of the positive torques due to low order $m$
corotation resonances, and perhaps because of mixing between the
torques from the CR and LRs of a given pattern (see discussion in
section \ref{sec:LRmagdiff} below). More accurate quantification of
the relative strength of the two sources of torques (from the CR and
the LRs), must be done in order to determine the true migration rate
of a planet. 

\subsubsection{The epicyclic frequency}\label{sec:epicycle}

In the theory of circumstellar disks, perhaps the two most critical
parameters describing the problem are the natural resonance frequency
of the system (the epicyclic frequency, $\kappa(r)$) and the driving
frequency (the planet's orbit frequency, $\Omega_{\rm pl}$). The
latter is very well determined since the planet is a single object.
The former is more difficult due to the important effects of pressure
and self gravity on the rotation profile. Nevertheless, a widely
implemented approximation in theory is that the differences from true
Keplerian orbits are small, so that the identity $\kappa(r)=\Omega(r)$
is approximately held. To what extent does the epicyclic frequency
deviate from equality with the orbital frequency? 

In figure \ref{fig:epicycle}, we show the ratio of the epicyclic
frequency at each location in the disk to the orbital frequency
$\Omega$ for the high resolution prototype simulations, with and
without self gravity. The variation of $\kappa$ is as large as 20\%
above and 10\% below the orbital frequency at each radius in the disk
for the self gravitating disk, but only $\sim$10\% in the non-self
gravitating disk. In the latter case, no gap is able to form due to
the very rapid migration of the planet. In the former, the largest
variation occurs about 2 Hill radii inward and outward from the
planet, which corresponds to the positions of local minima in the
surface density distribution in the nascent gap. Variation $>5$\%
extends to a distance of 2~AU inwards and outwards of the planet and
drops to a $\sim2$\% offset (due to the effects of pressure gradients
and self gravity on the rotation curve) at locations further inward
and outwards. Thus, although the equality between the two quantities
is not exact, differences are larger than 20\% nowhere in the disk and
are 10\% or less in most regions.

Differences of this magnitude remain important because as
\citet{Ward97} and others have pointed out, the effect on the
epicyclic frequency may be small but the effect on the gravitational
torque suffered by the planet can be very large. Modifying the
epicyclic frequency at each orbit radius will cause two separate
modifications of the torque. First, the resonance positions will
deviate from their Keplerian locations, and this change in position
will affect the value of the gravitational potentials calculated at
the resonance position. Second, the resonant denominator (for the LRs)
and the Oort constant, $B$, (for the CR) may change their values from
that predicted in an unperturbed disk. Are the differences in the
epicyclic frequency from the orbital frequency large enough to affect
strongly the values of these quantities? 

\subsubsection{The influence of the true rotation curve on the
resonance positions}\label{sec:reso-modpos}

In figure \ref{fig:resloc}, we show the resonance positions for all
three resonances (ILR, CR and OLR) as a function of $m$, for the high
resolution prototypes with and without self gravity. In only a few
cases, do the true resonance positions correspond to the ideal
Keplerian values. With self gravity, the CR positions are nearly
coincident with the planet's orbit, while without it, they shift
inward. At the same time, the LR positions are shifted due to the
additional terms in the definition of the resonant denominator. Large
$m$ resonance positions follow closely the buffer region (of size
$\sim 2H/3$) around the planet expected from the analysis of
\citet{art93a}. Without self gravity the inner resonances shift inward
by $\sim1$\%, but the outer resonances remain largely unaffected
because they are limited by the buffer region. For smaller $m$, the
positions follow the ideal Keplerian positions more closely, but
remain modified, especially for patterns $m\sim10$. For these
patterns, the resonance positions fall near the edges of the forming
gap, which means that they will be disproportionately affected by
large pressure gradients there.

If we follow instead the GT79 analysis and use the original definition
of $D$ of equation \ref{eq:D}, small deviations from the Keplerian
values are present both above and below the Keplerian values including
self gravity, but in general the correspondence is quite close.
Without self gravity, each of the resonances are systematically
shifted inward by about 1\% of the semi-major axis, which corresponds
to about 20\% of the Hill radius. The inner resonances are
systematically shifted further away from the planet, while the outer
resonances are shifted closer. Neither example displays the buffer
zone expected from the use of equations \ref{eq:Dstar} or
\ref{eq:Dsg}. 

\subsubsection{The influence of the true rotation curve on 
$|r dD/dr|_L$. }\label{sec:reso-modD}

The quantity, $D$, in equation \ref{eq:LR-flux}, defines the strength
and shape of the response of the disk to the perturbation from the
planet. In combination with the torque cutoff function (equation
\ref{eq:cutoff}), it determines the disk's response to the planet. In
the interests of mathematical tractability, it is usually approximated
by the first term in a Taylor series as $D\approx (r dD/dr)_{r_L} x$,
where $x=(r-r_L)/r_L$, an approximation that in turn is often further
approximated\footnote{We shall use the upper sign to refer to the ILR
and the lower sign to refer to the OLR} as $r dD/dr|_{r_L} \approx
-3(1\mp m)\Omega_{r_L}^2 \approx \pm3m\Omega_{r_L}^2$, which is
equivalent to the statement that the disk is Keplerian (i.e.
$\kappa=\Omega$). To what extent are these approximations valid after
the evolution begins? 

The quality of the approximation may be measured by the ratio of the
approximate value to the `real' value obtained numerically from our
simulations. A high value of the ratio means that more torques are
produced for that pattern in the simulation compared to what would be
predicted using the approximation. In figure \ref{fig:res-denom}, we
show this ratio for both the inner and outer Lindblad resonances. For
the self gravitating disk simulation, the approximation varies by up
to a factor $\sim2.3$ above and below the numerically obtained value
for patterns with $m\lesssim15$. Very sharp peaks exist in both the
OLR and ILR ratios, coming at $m=8$ and $m=10$ for the OLRs and ILRs
respectively. For the ILRs, the peak is broader and extends to $m=15$.
For lower $m$, the ratio drops below unity, indicating that the
simulation produces less torque than expected. Above $m=15$ and for
both the ILR and OLR ratios, the approximation and the simulation
produce very similar values, indicating that the approximation may be
used without large errors in the torque.

In contrast, the non-self gravitating disk model shows differences
only for the lowest $m$ patterns. For $m>10$ only relatively small
deviations of order a few percent are present and no large peaks are
present for any pattern although, as in the self gravitating version,
the ratio increases to as high as 1.3 and as low as 0.8 for the lowest
$m$ patterns. 

The patterns for which the approximation fails most severely (the
ratio peaks) are those most sensitive to the structure of a gap. Their
resonance positions tend to fall near the gap edges, both inside and
outside the planet. Since the planet is nearer to the inner edge of
the forming gap, more inner patterns, of higher order, are affected by
its presence. For higher order patterns, where the resonance position
falls well within the forming gap, the approximate form reproduces the
simulation value to within a few percent. For the non-self gravitating
simulation, the migration was so rapid that gap formation did not
occur. The resonant denominator was therefore not strongly affected
and the variations remained small. 

\subsubsection{The variation of $\Sigma/B$}\label{sec:CRquant-mod}

The strength of the CRs is proportional to the gradient of $\Sigma/B$
at the resonance, as indicated in equation \ref{eq:CR-flux}. This
quantity measures asymmetries in the circulation of matter as it
travels on horseshoe orbits in the planet's vicinity \citep{Ward91}. 
For a Keplerian disk, the Oort constant reduces to $B=\Omega/4\propto
r^{-3/2}$. Since our assumed surface density is also a power law with
the same proportionality, we expect that the value of $\Sigma/B$ will
be constant, and its gradient zero, making the CR torques
insignificant. Clearly, torques from near the CR exist in our
simulations. Is their origin due to the failure of this condition?

Figure \ref{fig:CRquant} shows the ratio of $\Sigma/B$ to its initial
value as a function of radius in the region near the CR. The slope of
this quantity presents a fair measure of the variation (i.e. the
gradient) in $\Sigma/B$ itself since the initial value is constant. In
the self gravitating simulation, $\Sigma/B$ decreases more than a
factor two below its initial value within $\sim2$\rh\ both interior to
and exterior to the planet, while remaining nearly unchanged at the
exact CR. The non self gravitating version displays variation of some
30\% below its initial value and over a wider radial extent. The
pattern is also offset relative to that seen with self gravity. While
CR appears near a local maximum of $\Sigma/B$ with self gravity, it
appears near a local minimum without it. We speculate that this fact
may simply be due to the very rapid migration in the latter case, so
that the planet simply outruns the disk's ability to keep up.

While at the exact CR, the slope may indeed be relatively small, we
recall the physical phenomenon of circulation on horseshoe orbits that
gives rise to the CR torques in the first place. In that context, the
conditions over the radial extent of the horseshoe orbit become
important, and using the exponential term in equation \ref{eq:CR-flux}
as our guide, we may associate the width of the resonance with the
disk scale height. In both simulations, the $\Sigma/B$ varies over a
radial scale of 2--4 disk scale heights and over this range, the size
of variation is as large as (or larger than) a factor of half its
magnitude and takes both positive and negative sign. We conclude that
evolution modifies the surface density and rotation profiles of the
disk enough to lead to very large gradients in $\Sigma/B$, and will
therefore also lead to significant CR torques.

\subsection{The sensitivity of the torques to the various physical and
numerical approximations}\label{sec:torq-sens}

Figures \ref{fig:epicycle}--\ref{fig:CRquant} show that variations
from the physical assumptions underlying the analytic derivations for
the gravitational torques do occur. Each of the quantities probes a
different aspect of the interaction, so that we can regard them as
acting independently and test the sensitivity of the torque to each in
turn. Which, if any, of the variations are important for correct
evaluation of the torque? Similarly, which, if any, of our numerical
assumptions are important?

\subsubsection{Torques at Lindblad resonances}\label{sec:LR-origin}

Figures \ref{fig:torq-mod} and \ref{fig:torq-mod-nosg} show the
torques as calculated under progressively less exact correspondence to
equation \ref{eq:LR-flux}, in which more precise quantities are
cumulatively replaced with commonly used approximations. Significant
differences in the one sided torque magnitudes occur when moving from
nearly any panel to any other in either figure. The magnitude up or
down of these variations is as much as a factor of several depending
on the approximation. Since two one sided torques (each of opposite
sign) must be added together, they produce a small net torque that can
vary by a factor 3--10 and that can even change the sign of the torque
for some patterns.

Even though the disk is highly stable against self gravitating
instabilities (i.e. $Q>5$ everywhere), the difference between the (a)
and (b) panels of figure \ref{fig:torq-mod} shows that disk self
gravity remains an important effect through the effective resonant
denominator $D_{sg}$. A $\sim20-30$\% reduction in the calculated
torques occurs when the self gravity term is suppressed (i.e. $D_*$ is
used instead). This difference occurs mainly because the positions of
each resonance are shifted farther from the planet in the latter case.
A second difference due to self gravity is that rotation curve itself
shifts, so that the inner and outer resonances will be systematically
shifted with respect to the planet (figure \ref{fig:resloc}). The
effect of this difference is clear when we examine the relative
magnitudes of the torques from the (b) panels of figures
\ref{fig:torq-mod} and \ref{fig:torq-mod-nosg}. The outer (negative)
torques on the planet are systematically larger in magnitude (more
negative) without self gravity than with it.

The one sided torques for higher $m$ patterns increase by 20\% (or
more for the highest $m$'s) in panel (c), relative to (b), where we
substitute the torque cutoff function and $D$ for the generalized
Laplace coefficients and $D_*$. The physical interpretation of this
replacement lies in how the three dimensional disk potential (implied in
the generalized Laplace coefficients) and non-WKB terms are accounted
for in the derivation. The change in one sided torques leads to a
factor $\sim3$ increase in the calculated net torque on the planet,
which is quite consistent with the analytic discussion of the effect
predicted for this replacement discussed by \citet[][see esp. his
figure 1]{art93b}. Together, the generalized Laplace coefficients and
the resonance buffer region act very effectively to suppress the
torques for high $m$ patterns. 

For the non self gravitating version, another change is much more
dramatic. The outer torques (especially for higher $m$) increase in
magnitude by as much as a factor 10, while the inner torques remain
essentially unchanged. The primary culprit in this case is the
resonance position shifts, due to the replacement of $D_*$ with $D$ as
resonant denominator. When the buffer region produced using $D_*$ is
suppressed, both the inner and outer resonances are shifted inward
relative to those in the self gravitating disk, with the outer
resonances being closer to the planet (producing a larger torque). The
inner resonances are shifted inward relative to the Keplerian
positions, but end up being shifted nearly to what were their buffered
positions, and are therefore not greatly affected.

If we account exactly or approximately for local variation in the
rotation profile (i.e. variation in $|r dD/dr|_{r_L}$ from figure
\ref{fig:res-denom}), we find large changes in the one sided torques
for patterns with $m\approx 7-12$, due to the small scale rotation
curve variation near the forming gap edges. When we replace the true
(first order approximation of the) resonant denominator with its
approximate form, $-3(1\mp m)\Omega_{r_L}^2$, in panel (d), the peaks
become much less significant. The torques in figure
\ref{fig:torq-mod-nosg}(d) show no features comparable to those in
figure \ref{fig:torq-mod}(d), because gap formation is much less
advanced and little variation in the resonant denominator exists.
Therefore we conclude that self gravity plays only an indirect role in
the variation of the resonant denominator, through the effect the
migration rate has on the disk profile.

Even though $m$ to $m$ variations are present in the torques from the
simulations, no variations of similar magnitude to those from the
analytic calculations shown in figure \ref{fig:torq-mod}(a)--(c) are
present. The answer to this apparent paradox becomes clearer upon
examination of figure \ref{fig:cum-torq}. The response of the disk
(i.e. the torques) occurs over about 1/2 cycle of the wave rather than
at a single radius, and the shape of the response varies with each
pattern. We conclude that in order to account for this spatial
variation, either an average which adequately accounts for the
variation or terms of higher order than the first order approximation
currently implemented must be used. Perhaps surprisingly, the
approximation $|r dD/dr|_{r_L}=-3(1\mp m)\Omega_{r_L}^2$ appears to be
such an average, at least for the patterns affected most by gap
formation ($m\approx10$). The same conclusions may remain somewhat
useful for use in conjunction with the modified resonant denominators,
$D_{sg}$ or $D_*$ as well, in spite of the fact that use of this
approximation is only justified in combination with the torque cutoff
function, because the differences are small when $m$ itself is not too
large (specifically for the patterns affected most by the gap edge
pressure gradient effects). Due to space limitations, we have not
discussed another common approximation of the resonant denominator, as
$r dD/dr|_{r_L}=\mp 3m\Omega_{L}^2$. However, we note that this
approximation also leads to net torque changes of a similar magnitude
to the others in panels (d)-(f). 

Accurate or approximate determination of the resonance positions
provides an additional significant source of variation in the
calculated torques. Figures \ref{fig:torq-mod}(e) and (f) shows two
approximations of the resonance positions. Panels (e) show the effect
of replacing the position determined by the `zero' of the resonant
denominator with exact $(m\pm1)/m$ ratios to the orbit frequency.
Changes of almost a factor two occur in the self gravitating disk net
torques. In the non self gravitating disk only small differences
occur, presumably due to the already very large shifts from other
sources. This replacement is equivalent to saying that $\kappa=\Omega$
everywhere, which we have shown in section \ref{sec:epicycle} to be
accurate to 20\%. In the (f) panels, we show effect of another
approximation of the resonance positions, namely that the argument of
the Bessel functions (see equation \ref{eq:Lap-Bes}), $m(1-r_L/a)$, is
exactly 2/3. While the torques are affected only at the $<5$\% level
by the Bessel function approximation itself, a change of order unity
is seen in the resulting torques on the planet when its argument is
approximated. 

Based on the changes we observe when we implement various
approximations, we can make several important conclusions. First, and
most importantly, the torque calculations are exquisitely sensitive to
nearly all of the common approximations used to calculate them.
Qualitatively and quantitatively incorrect results will be obtained
unless rigid adherence to the mathematically derived formulation is
strictly observed, and a physically inclusive model is developed.
Second, a theoretical model specifically including disk self gravity
is important for correct evaluation of the torque, even in disks where
self gravitating spiral density waves are largely suppressed (i.e.
$Q>>1$ everywhere) because it acts to shift the resonance positions.
Third, use of the torque cutoff function may be completely avoided
when both the resonant denominator is modified according to equation
\ref{eq:Dsg} and the Laplace coefficients are modified according to
equation \ref{eq:Lap-vert}. We would caution that such a replacement
may be of limited value however, since the true vertical structure of
the disk (modeled with the generalized Laplace coefficients) remains
an approximation.

\subsubsection{The LR torque magnitude differences, and an attempt to
obtain correspondence}\label{sec:LRmagdiff}

Even when each of the approximations in section \ref{sec:LR-origin}
are accounted for, the predicted torques do not reproduce those
obtained from the simulations, which we saw were a factor 3--6 smaller
than predicted. In very recent work, \citet{DHK02} have also reported
a similar discrepancy between analytic predictions and simulations,
including for planet masses far below those in our own study, for
which perturbations are very small and simulations are very far from
any possible non-linear effects. They implement a nested grid method
to obtain extremely high resolution of the mass in two dimensions
around the planet, and consider a gravitational softening smaller than
ours. What is the origin for this large difference, occurring in both
results?

It is possible to eliminate the effects of both finite grid resolution
and gravitational softening as a cause for the differences.
\citet{DHK02} provide extremely high grid resolution near the planet
and experience the same phenomenon, while gravitational softening is
accounted for theoretically through our implementation of the
generalized Laplace coefficients from equation \ref{eq:Lap-vert}.
Instead, we suggest that the origin of the differences may be another
consequence of the finite width of the resonances (recall also the
discussion of the resonant denominator, in section
\ref{sec:LR-origin}, above).

In a derivation resulting in the same torque formulae for LRs
originally developed by GT79, \citet[][hereafter MVS]{MVS87} define
expressions for the resonance width in a disk with pressure, self
gravity or viscous forces (or any combination) acting on the fluid. As
a condition of validity, they point out that the resonance must be
narrow compared to the distance between the resonance and the planet
(in non-dimensional form $w<<1/m$, with $wa_{\rm pl}$ the true width)
so that the orbits of the planet and resonating material do not cross.
This condition is equivalent to the tight winding approximation used
in the torque derivation. They suggest that failure of this criterion
will lead to mixing of the torques from inner and outer resonances and
a decrease in the torque exerted. 

MVS derive a width for a Lindblad resonance of
\begin{equation}\label{eq:mvs-pwidth}
|w|^3  = {{ c^2/\Omega_{r_L}} \over { 3m r_L^2 \Omega_{\rm pl}}}
\end{equation}
assuming that pressure dominates over self gravity and viscosity.
Substituting values for the variables appropriate for our simulations,
we find that $w\approx0.1m^{-1/3}$. For $m=10$ and $m=20$, we thus
obtain $w\approx0.045$ and $w\approx0.037$ respectively, which is
comparable to the distance between the planet and the resonance and
suggests that some overlap and mixing may occur for these patterns.
Further, if mixing between the LRs and CR can occur (as we discuss in
the following section), then the amount of overlap will be larger
since the distances are smaller.

There are several important caveats to consider before this conclusion
can be substantiated however. First is the fact that the pressure wave
propagates {\it away} from CR, and so has very low amplitude close to
the planet available for mixing. The self gravity term will be
responsible for a second wave in the disk which propagates towards the
CR. Its width (from the MVS formulation) is
$w\sim7\times10^{-2}m^{-1/2}$ in our simulations, nearly as large as
the pressure term. It is interesting to speculate that this term could
also be active, though there is no direct evidence for it.

Secondly, as discussed in section \ref{sec:reso-modpos}, the LRs will
shift due the addition of non-WKB terms, neglected in the MVS
analysis. Indeed, the entire derivation of MVS is superseded by the
later \citet{art93a} analysis, in which the issue of the resonance
widths was not specifically addressed. Close examination of figure
\ref{fig:resloc} shows that the buffer zone does not begin to modify
the resonance positions until $m\approx20$, so if we naively use the
resonance widths from the WKB theory with the resonance positions from
the non-WKB theory, we still may find significant overlap, though less
than might originally be expected from the MVS analysis.

It is beyond the scope of this work to investigate further the
possibility that the differences between theory and simulation are due
to resonance overlap or to some other effect. However, it may still be
useful to obtain information about the character of the differences
for future investigations. We saw in \jone\ that the simulations were
in fact very sensitive to the softening implemented in the simulation.
Here we ask a similar question about the torques from the analytic
derivations. We attempt to obtain the best correspondence possible
between the simulation and theoretical torques, by adjusting the
available parameters in the analytic theory. 

Figure \ref{fig:torq-soft} shows the analytic predictions torques for
the self gravitating simulation, assuming a gravitational softening
factor 2.5 larger than was actually present in the simulation itself.
We also use the approximated resonant denominator, $-3(1\mp
m)\Omega_L^2$ for patterns with $m\le15$, but retain its numerically
derived value for higher $m$ patterns. In both the self gravitating
and non self gravitating versions, the inner (positive) torques are
reproduced much more faithfully than the outer (negative) torques
especially without disk self gravity. Overall, except for very high
$m$ patterns, where the analytic torques underestimate those from the
simulations due to the increased softening, the agreement is within
$\sim20$\% for any given pattern in the self gravitating simulation.

It is an encouraging sign for the theory that we are able to reproduce
the behavior of the $m\le15$ patterns, since those are the ones for
which the resonant denominator is most strongly affected by the
forming gap and for which the variation in softening will have the
least effect. However, the non self gravitating version fares much
worse, with large differences for nearly all patterns. Thus these
after the fact parameter adjustments should carry little weight.

\subsubsection{Torques at Corotation}\label{sec:CR-origin}

As we noted above, the torques due to interactions at the CR are
characterized at least as well by their differences from simulation to
simulation than by their similarities. Except to say that they are
significantly nonzero, we were unable to determine accurately their
magnitude due to our limited spatial resolution. What conclusions can
be made about the importance of the interactions at corotation based
on the causes of these variations in our simulations?

Both with and without disk self gravity, the CR position is very close
to the planet, making quite large torques possible. In fact, the
position in the self gravitating disk version is very nearly
coincident with the planet's orbit radius. With self gravity
suppressed it lies further inward. In spite of its greater distance,
the CR torques are larger in the non self gravitating versions, so the
separation between the CR and the planet is not by itself the
determining factor in the torque magnitude.

\begin{figure}
\psfig{file=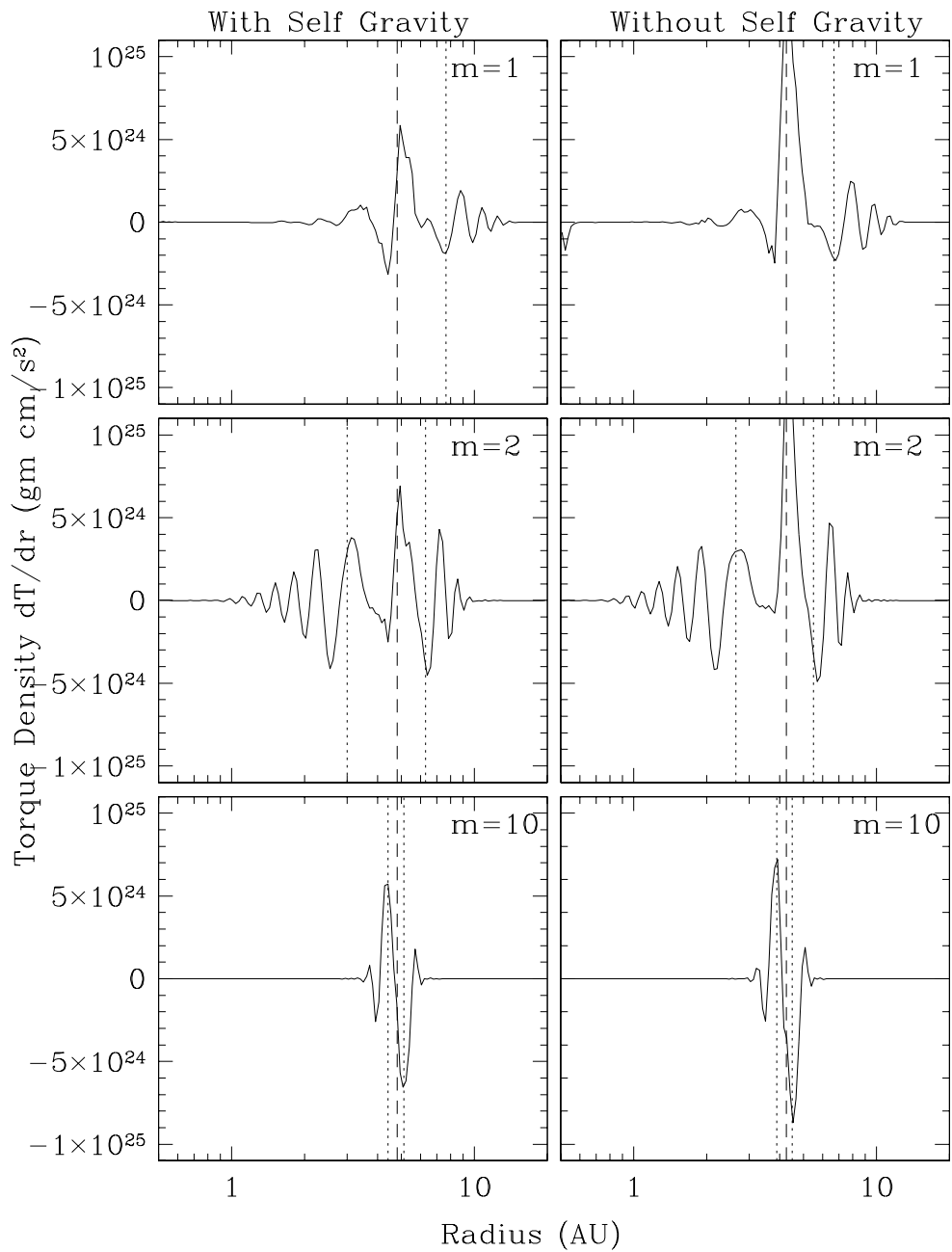,height=7.2in,rheight=6.9in}
\caption[Radius dependent torques for the low mass prototype]
{\label{fig:torqradm-lo}
Torque density of the disk on the planet as a function of distance
from the star, for the low mass prototype simulation, with (left) and
without (right) disk self gravity. The three panels from top to bottom
show the contributions to the torque derived from spiral patterns of
$m=1,2$ and 10 symmetry, as marked. The orbit radius of the planet is
shown with a vertical dashed line in each panel, and the Lindblad
resonances with vertical dotted lines. Large torques very close to the
planet are due to the corotation resonances. }
\end{figure}

\begin{figure}
\psfig{file=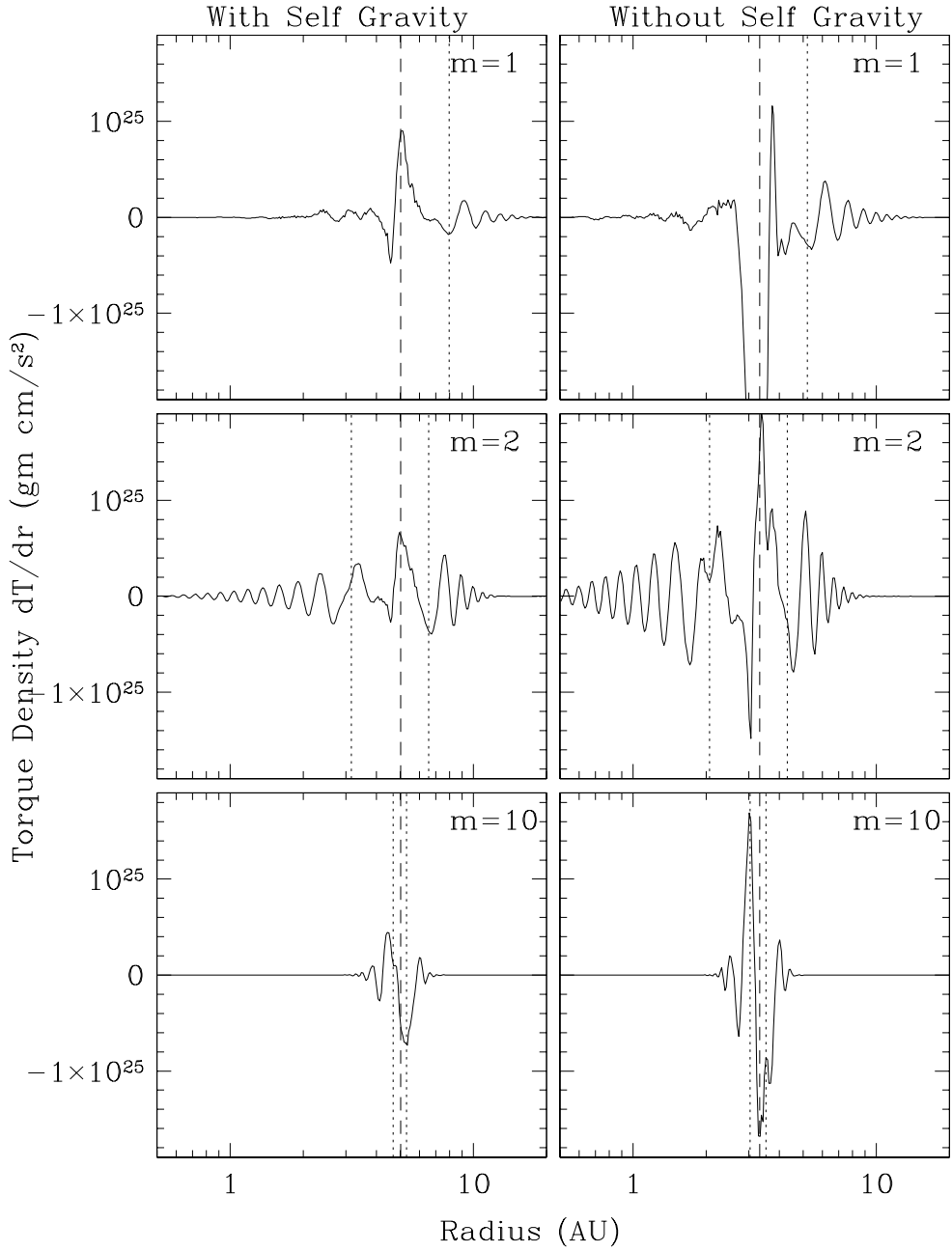,height=7.2in,rheight=6.9in}
\caption[Radius dependent torques for the high resolution prototype]
{\label{fig:torqradm-hires}
Like figure \ref{fig:torqradm-lo}, except for the high resolution
prototype simulations. }
\end{figure}

\begin{figure}
\psfig{file=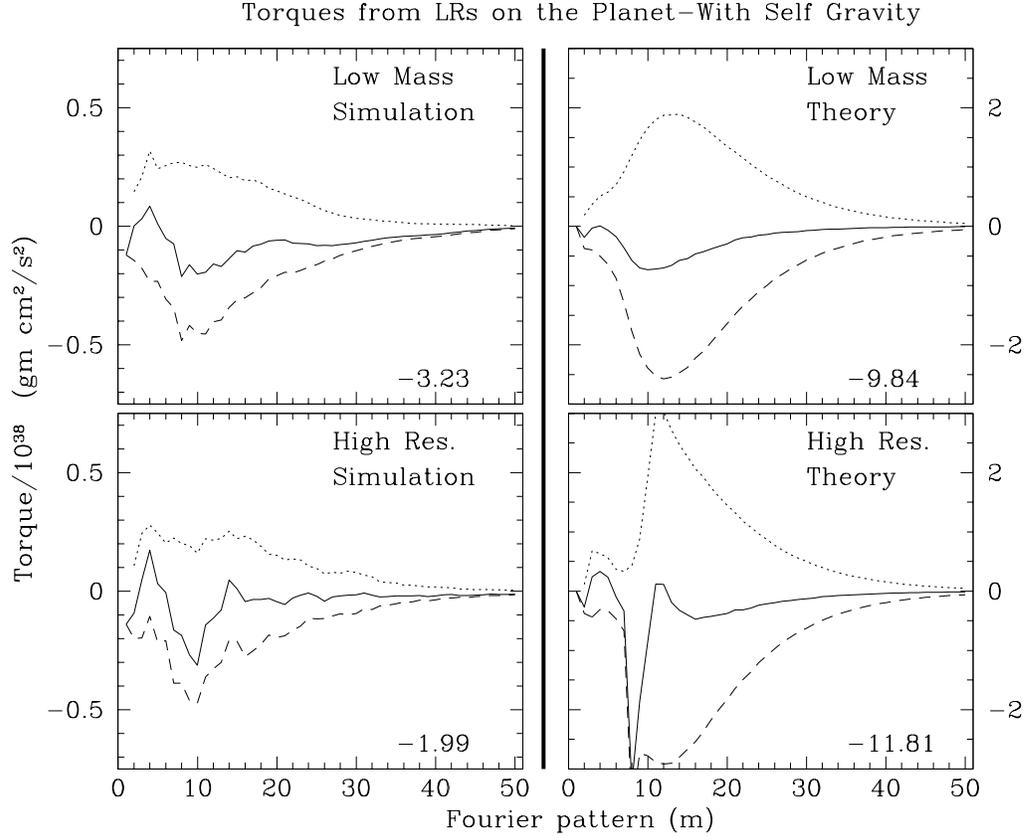,angle=-90,height=4.5in}
\caption[Mode torques]
{\label{fig:mtorq-LR}
Torques of the disk on the planet generated by the Lindblad resonances
with spiral symmetry $m$ ($m\le50$) on the planet at time, $t=300$~yr. As
indicated, the top and bottom panels show the torques as calculated for
the low mass and the high resolution prototype models. In each case, the
total torque is shown with a solid line, while the torques from each ILR
and OLR are shown with dotted and dashed lines, respectively. Note that the
axes are expanded by a factor of four in the theory plots relative to the
simulations. The number in the lower right corner of each panel is the 
net torque from the sum of all LRs. } 
\end{figure}

\begin{figure}
\psfig{file=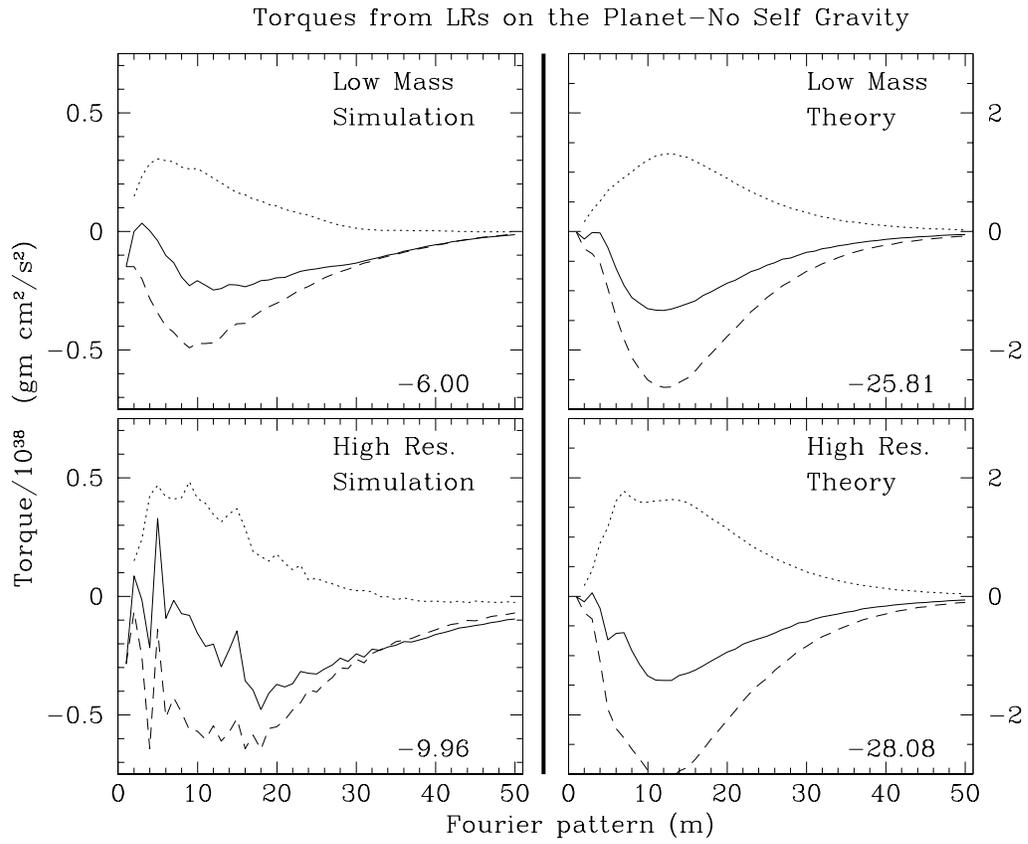,angle=-90,height=4.5in}
\caption[Mode torques]
{\label{fig:mtorq-LR-nosg}
Same as figure \ref{fig:mtorq-LR}, except for the simulations without
disk self gravity. } 
\end{figure}

\begin{figure}
\psfig{file=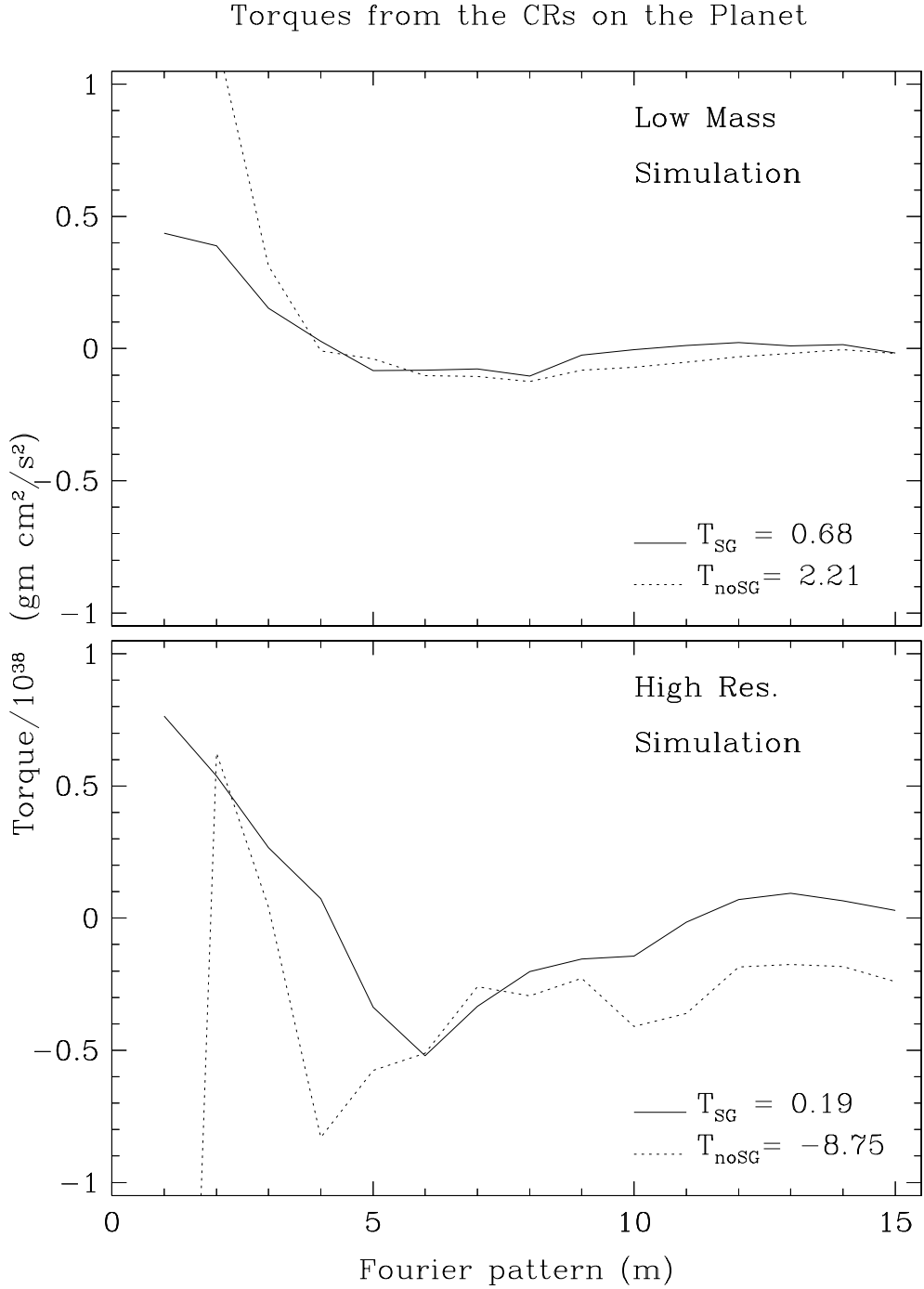,angle=0,height=7.5in}
\caption[Mode torques]
{\label{fig:mtorq-CR}
Total torques of the disk on the planet generated by corotation resonances
with spiral symmetry $m$ ($m\le15$), for the low mass prototype and the high
resolution prototype with (solid line) and without (dotted line) disk self
gravity. The net torques from the CRs are shown in the lower right corner.} 
\end{figure}

\begin{figure}
\psfig{file=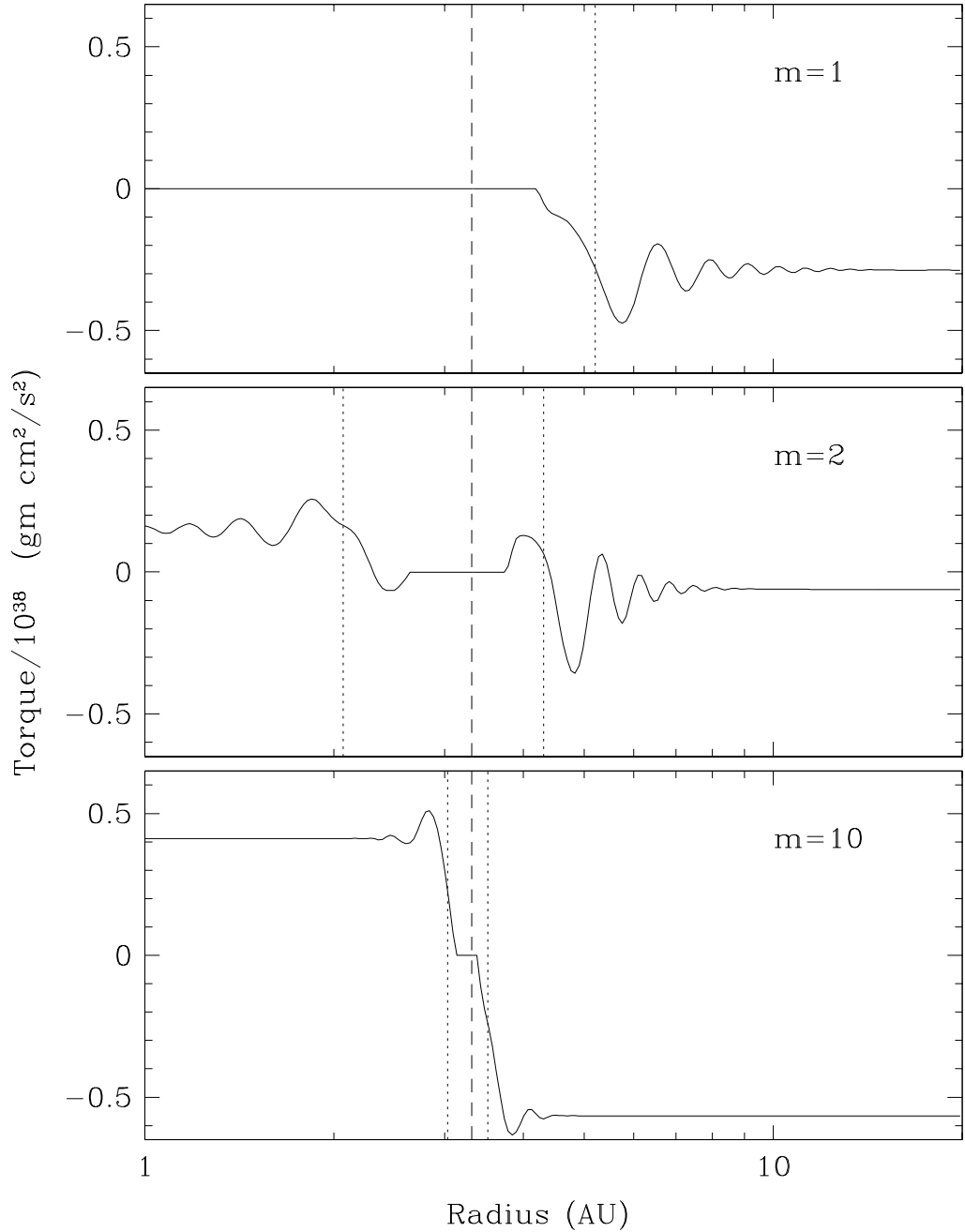,angle=0,height=7.5in,rheight=7.3in}
\caption[Cumulative Lindblad torques]
{\label{fig:cum-torq} 
The cumulative torque from the inner and outer Lindblad resonances
due to the $m=1$, 2 and 10 patterns in the high resolution prototype
simulation (without self gravity). The LR positions derived from the
simulation are marked with vertical dotted lines and the planet's
position is marked with a dashed line. The cumulative sum is taken as
a function of distance from the planet for both the inner and outer
resonances. Infinite distance inside or outside the planet's position
yields the net torque from that resonance. } 
\end{figure}

\begin{figure}
\psfig{file=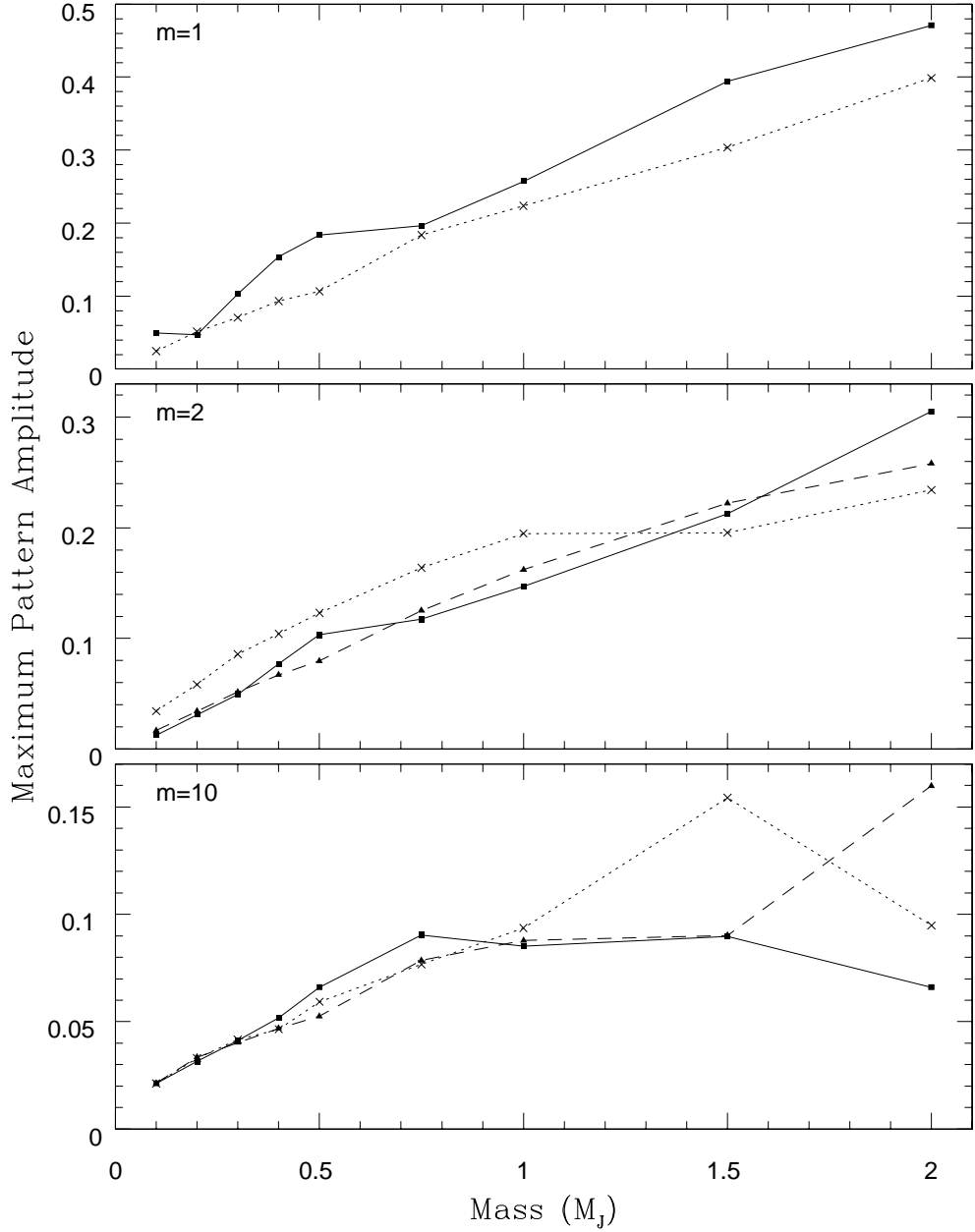,angle=0,height=7.5in,rheight=7.0in}
\caption[Pattern amplitudes]
{\label{fig:maxpatamp}
Maximum Pattern amplitudes vs. planet mass for the $m=1,2$ and 10
patterns between 0.1 and 2.0\mj. In each plot the solid line with
filled squares represents the amplitude at corotation, the dotted line
with crosses ($\times$) represent the amplitude at the OLR and the
dashed lines with the filled triangles represents the amplitude at the
ILR. In all cases, the amplitude is normalized to the azimuth averaged
surface density at the appropriate resonance.}
\end{figure}

\begin{figure}
\psfig{file=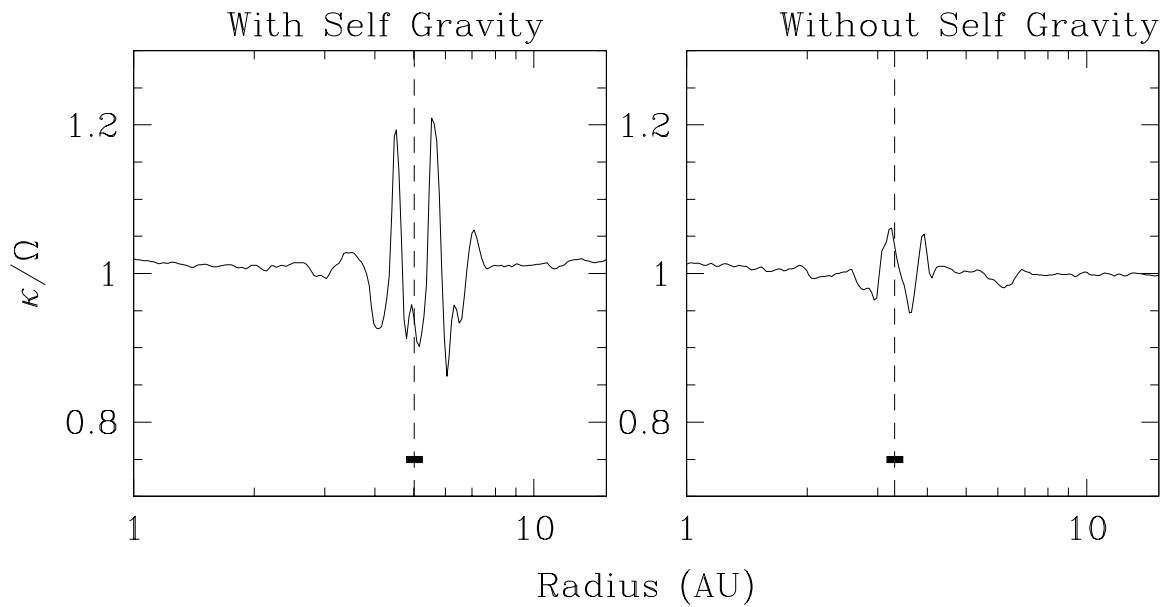,angle=0,height=8.5in,rheight=3.3in}
\caption[Epicycle]
{\label{fig:epicycle} 
Ratio of the epicyclic frequency, $\kappa$, to the orbital frequency,
$\Omega$, at each radius for the high resolution prototype with and
without self gravity, both at time $t=300$~yr. In the gap formation
region ($a_{\rm pl}\pm4$\rh) the quantities differ by as much as 20\%
due to the influence of large pressure gradients on the rotation
curve. Outside the gap formation region, the ratio is near 1.02,
indicating the expected near equality of the two quantities. Migration
is extremely rapid in the non-self gravitating system and consequently
no gap is able to form and the ratio stays near unity. In each panel,
the vertical dashed line defines the position of the planet, and the
thick horizontal bar defines one Hill radius around the planet.}
\end{figure}

\begin{figure}
\psfig{file=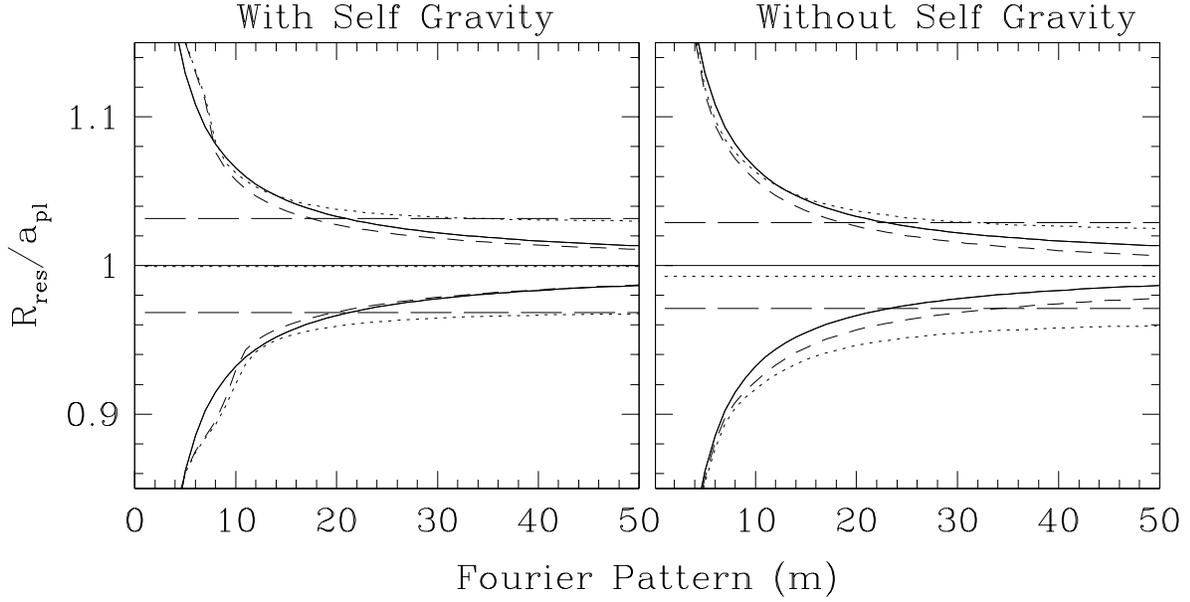,angle=0,height=8.5in,rheight=4.0in}
\caption[Resonance location variation]
{\label{fig:resloc}
Location of the Lindblad and corotation resonances as determined from
our simulations. The left and right panels show the resonance
positions for the simulation with and without self gravity,
respectively. The ideal Keplerian resonance positions are shown with
solid lines, while the true positions (calculated using equation
\ref{eq:Dstar} or \ref{eq:Dsg}) are shown with dotted lines and the
positions calculated from equation \ref{eq:D} are shown with short
dashed lines. The horizontal long dashed lines define the extent of
the buffer regions ($2H/3$) inside and outside the planet, from
\citet{art93a}.} 
\end{figure}

\begin{figure}
\psfig{file=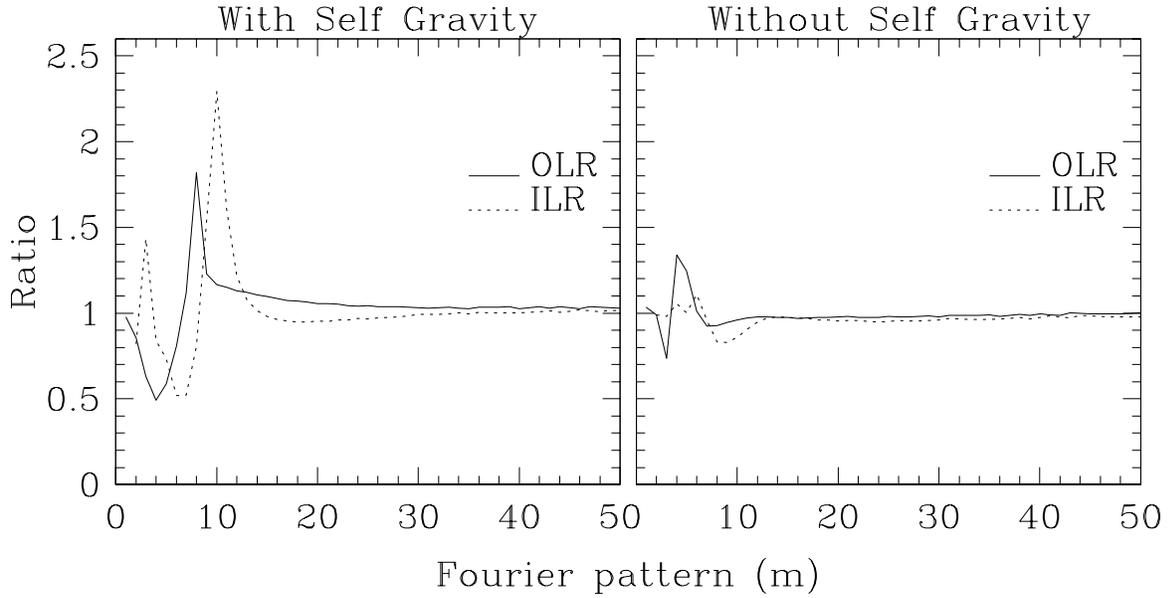,angle=0,height=8.5in,rheight=4.0in}
\caption[Resonant denominator]
{\label{fig:res-denom}
The ratio of the approximated value of the resonant denominator to the
value computed numerically from the high resolution prototype
simulations, with and without self gravity. The ratio is defined as
$R= -3(1\mp m)\Omega_{r_L}^2 / (r dD/dr)_{num}$. 
As marked, the ratios at the inner resonance are shown with dotted
lines while the outer resonance ratios are shown with a solid line. } 
\end{figure}

\begin{figure}
\psfig{file=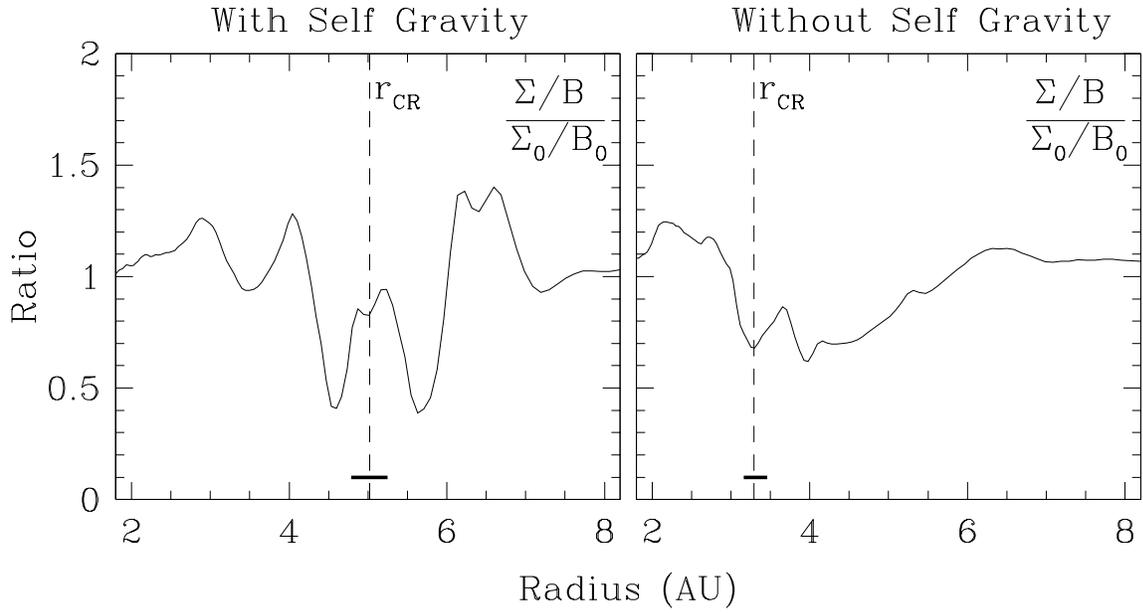,angle=0,height=8.5in,rheight=3.5in}
\caption[Derivative of the CR sensitive parameter Sig/B]
{\label{fig:CRquant}
The ratio of the CR parameter $\Sigma/B$ to its (near constant)
initial value for the high resolution prototype simulations. The
solid bar near the bottom of the plot, defines the size of the Hill
sphere.}
\end{figure}

The torque produced at corotation is exerted within a distance of a
few scale disk heights of the resonance (see the exponential term in
equation \ref{eq:CR-flux}), but its theoretical magnitude is
determined only at one point in that range--the exact resonance
position. Therefore, while the analytic derivations account for the
width of the resonance they do not account for large variations in the
disk quantities over the spatial scale where the torques are active.
Variations in these quantities (e.g. $\Sigma/B$) of order unity are
present over these same few disk scale heights. Further, as
\citet{KP93} also point out in a much different analysis, the planet's
gravitational potential may also vary significantly over the same
spatial scale. Both issues will render attempts to compare to theory
of limited use, and we conclude that the approximation that the
conditions only at the exact resonance define the torque cannot be
supported for corotation resonances. Figures \ref{fig:torqradm-lo} and
\ref{fig:torqradm-hires} showed that both sign changes and large
magnitude changes in the torque do occur over this same range,
strengthening this conclusion. 

In some sense, the quantity $\Sigma/B$ is a measure of the flow
pattern of mass around the planet. \citet{LSA99} studied the mass flow
around the planet during the \ttwo\  migration period at a factor
$\sim$20 higher resolution than our simulations and showed that mass
transfer through the gap can occur efficiently. They also find that a
circumplanetary disk develops and a large amplitude, double armed
spiral pattern forms within it. The mass transfer they see will
inevitably lead to a net torque on the planet, and as they discuss,
the net torque from the small radial region around the planet is as
large as that from the rest of the disk. Presumably, the contribution
during \tone\ migration will only be larger, making CR torques even
more important for the migration than they indicate. However, the
conditions in their work are dissimilar from our own in a very
significant way--we showed in section \ref{sec:plan-env} that a
circumplanetary disk was very unlikely during \tone\  migration. How
much will the flow pattern and torques during the \tone\  period
change due to the existence of a (properly resolved) rotating envelope
structure rather than a disk? 

Although we quite naturally expect CR torque to be sensitive to the
$r$ coordinate, sensitivity in the angular coordinate is missing in
equation \ref{eq:CR-flux}. In separate analyses
\citep{Ward91,M01,BK01}, the sensitivity of the CR torque to the
angular coordinate is recovered in the context of angular momentum
exchange between the planet and disk material in the horseshoe region
of the disk. However, these authors also conclude that the CR torques
will not in the end be very important because the circulation around
the horseshoe orbits will reach a steady state in a relatively short
time and so the torques will decrease to zero. Is the fact that we
start without a saturated circulation region the reason that the CR
torques are large in our models?

A possible `escape route' for the saturation argument has been that if
the planet migrates, then the region affected by the CR torques will
continually change and the saturation will never occur. While this may
still be true, we believe there is an equally plausible alternative
mechanism--namely the changes in the disk's radial profile caused by
the actions of the Lindblad resonances. The question is relevant
because as we concluded above, the CRs act over a finite region (of a
few disk scale heights) that overlaps with the higher order $m$ LR
locations. If the LRs can produce changes in the disk profile on a
similar or shorter time scale to the one at which CR interactions
smooth them out, then saturation will never occur. \citet[][see esp.
their figure 8]{BK01} showed that the time scale for saturation is
\begin{equation}\label{eq:tauCR}
\tau_{\rm CR}\approx {{5P}\over{2\pi}} \sqrt{{{M_\odot}\over{M_{\rm pl}}}},
\end{equation}
where $P$ is the planet's orbit period. A relevant quantity for the
LRs is the gap formation time scale \citep{Bryden99} 
\begin{equation}\label{eq:tauLR}
\tau_{gap}= P \left({{M_\odot}\over{M_{\rm pl}}}\right)^2
                    \left({{\Delta}\over{a_{\rm pl}}} \right)^5 
        =  3^{-5/3}P \left({{M_\odot}\over{M_{\rm pl}}}\right)^{1/3}
\end{equation}
where the right hand equality is obtained from setting the gap width,
$\Delta$, equal to the Hill radius, where most of the LR positions are
found. Relating these two quantities to each other yields
\begin{equation}\label{eq:tau-rat}
\tau_{\rm CR} =  3^{5/3} {{5}\over{2\pi}}
              \left({{M_\odot}\over{M_{\rm pl}}}\right)^{1/6}
                              \tau_{gap}.
\end{equation}
The saturation time scale for the CRs is thus comparable to or larger
than the gap opening time scale for Jovian and sub-Jovian mass planets
and the CRs will always be active.

\section{Discussion and Comparisons to Other Work}\label{sec:summary}

In this section we attempt to compare our results with some of what
has gone before, to place them in context. Many previous numerical
studies of disks with embedded planets have also included mass
accretion onto the forming planet. Some have begun, as we do, with an
unperturbed disk \citep{Bryden99,NPMK}, while others begin with an
already formed gap \citep{LSA99,Kley99,KDH}. One study concentrates on
very high resolution local simulations of a small region of the disk
containing the planet \citep{Miyoshi99}. In many instances these
studies have also suppressed migration in order to explore the gap
formation and accretion processes in isolation. 

Similar to \citet{Bryden99} and \citet{LSA99}, we find that matter
accretes primarily from inside and outside the planet's orbit, rather
than from the region radially closest to the planet. The planet masses
reached at the end of our accretion simulations are similar to those
of found by others, but the accretion rates in our simulations are
much larger than those of the previous work. The differences in the
time scales are dominated by the differences in assumptions about the
initial conditions--they start with a nearly empty gap region, while
we do not. Secondarily, their simulations assume that accreted mass
does not accumulate on the planet--its mass remains constant--since
they were primarily interested in studying the properties of the flow
through the gap. Common to both our models and theirs is that the
planet accretion is self limiting. Planet apparently cannot accrete
more than $\sim5-10$\mj\  of mass from the disk starting either from
the initial condition of an unperturbed disk or from one with an
already formed gap, because additional mass is driven away by
dynamical torques. 

This conclusion is interesting on both theoretical and observational
grounds. Specifically, Jovian planet growth was long thought to be a
self limiting process due to gap formation \citep{PP3_LP}, but
recently had been challenged by the possibility of mass accretion
through the gap \citep{Bryden99,Kley99,LSA99}. Our combined results
show that while growth and accretion continues with or without a gap,
there is a clear limit to this process, no matter what initial
conditions (gap/no gap) are used. On the observational side, the
masses reached by our simulations and Kley's are interesting because
they are comparable to the largest `minimum masses' (i.e. $M\sin{i}$)
observed for low mass companions in short period orbits around other
stars.

Our results for the sign of the accretion torque are opposite those of
\citet{NPMK}, who show that accreted matter brings with it net
positive angular momentum. Our work shows net negative angular
momentum accretion. While our conclusion and theirs are the same--that
the influence of accretion torques on the migration are small--the
difference in sign of the torque is troubling. There are a number of
differences in the disk morphologies, which could affect the result.
In particular, the surface density distribution we use is steeper than
theirs, going as $r^{-3/2}$ rather than the flat, $r^0$ profile they
use. It seems quite plausible that with such distribution the amount
of matter accreted from outside the planet's orbit (i.e. with greater
angular momentum) will be higher than with a steeply decreasing
profile, causing the sign of the net accreted torque to be positive.

Our {\it acc0} and {\it acc1} simulations (with accretion factor,
$f=1.0$ and $0.1$, respectively) have very similar initial conditions
and physical assumptions to those discussed in \citet{AH99}, except
that their disks were about twice as massive as ours, and their
interest was on triggering further planet formation rather than
migration. Nevertheless, the outcomes are fundamentally quite
different. They find that rapid growth of the planet (from 1\mj\ to
4-5\mj\  in less than 30~yr) ends in fragmentation of the growing
spiral structure into additional planetary mass companions in fewer
than 3 orbits of the original planet. While we also find rapid spiral
structure growth as the planet gains mass and perturbs the disk, we do
not find any signs of the fragmentation they observe. In work in
preparation (Nelson 2003), we show that their fragmentation is instead
numerically induced by a violation of a two dimensional version of the
\citet{Truelove97} criterion applicable in disks. This condition is
satisfied in our simulations, which do not fragment. 

In general, each of the numerical works above focus their attention on
the phenomenological aspects of the evolution (gap formation,
accretion processes) and spend much less effort in direct comparisons
with theory. In our work, we have made a detailed comparison of the
results of our simulations and the theoretical predictions for both
Lindblad and corotation interactions. We have attempted to examine the
assumptions underlying the development of the analytic torque
formulae, that are made in order to make the problem tractable enough
to solve mathematically. The theoretical work to which we compare is
based on the seminal studies of GT79, GT80 and \citet{LP79} and have
in general focused either on effects that determine the Lindblad
torques \citep{LP86,MVS87,Ward86,art93a} and the formation of the gap
\citep{Takeuchi96,Ward97}, or on the character of the corotation torques
\citet{Ward88,Ward91,M01,BK01}. 

Many of the qualitative predictions of the theories are reproduced in
our simulations. We find that dynamical torques develop exhibit
wavelike radial structure that decays with distance from the planet in
the regions dominated by Lindblad interactions, and non-wavelike
structure close to the planet where corotation resonances are strong.
Since the Lindblad interactions are sums of two terms from the inner
and outer resonances (except $m=1$) with opposite sign, the net
contribution from any given spiral wave pattern, $m$, is much smaller
than from any single resonance considered in isolation.

When we look in more detail however, correspondence is weaker. The
expected linear dependence of the migration rate on planet mass is not
reproduced (see \jone) and the torque magnitudes from Lindblad
resonances obtained from the simulations are a factor 3-6 smaller than
those predicted from theory. Secondly, specific features of the
torques as a function of pattern number $m$ do not correspond well in
most approximations, especially for patterns whose resonances fall
near the edges of the forming gap. 

Two of the most critical assumptions made by GT79 are that the
interactions produce perturbations in the disk that are small so that
linear theory may be applied, and that all interactions between the
disk and planet occur exactly at either a Lindblad or corotation
resonance. We have checked both of these assumptions in our
simulations and have found that they can be supported, at least in
part, for planets with mass \mplan$\lesssim0.5$\mj. For these models,
the perturbation amplitudes at each resonance are proportional to the
planet mass, which agrees with the assumption that the perturbations
in the disk should be linearly proportional to the size of the
perturber. For higher mass planets, the pattern amplitudes of the most
important patterns ($m\sim10-20$) become saturated and no longer vary
as a function of planet mass. Computations that depend on the validity
of the torque formulae for these masses may suffer from inaccuracy.

While it is true that some torque is exerted far from the resonances
in our simulations (and indeed, is expected--see e.g. \citet{Ward86}),
we do find that by far the largest fraction of the torque from any
given pattern is exerted over the first cycle of the wave at a
Lindblad resonance. This validates in part the analytic derivation of
the angular momentum flux discussed in GT79, which depended on the
planet's potential varying little near the resonance and the tight
winding assumption. A closer examination however, shows that we cannot
completely validate the mathematical derivation in our simulations
because the width of the resonance is not negligible, and the shape of
torque response within that resonant width varies from one pattern to
the next, making a determination of the true effective resonance
position difficult. 

This variation is especially important in the context of approximations
of those positions made to ease calculation of the torques. We have
shown that small shifts in the resonance positions can have a very
large effect on the calculation of the torque on the planet. The fact
that shifts due to pressure gradients could produce large changes in
the torque is well known \citep{art93a,Ward97}, but we are not aware of
any work (analytic or numerical) that considers the shifts due to disk
self gravity. Our results indicate that the shifts due to self gravity
are approximately the same magnitude as those caused by pressure
gradients, but opposite in direction. Migration rates are much slower
including disk self gravity than without it. 

Of particular interest in our torque comparison is the fact that
torques from the simulations are a factor of 3--6 smaller in magnitude
than those from analytic models that include the effects of vertical
structure and the buffer region around the planet, as outlined by
\citet{KP93,art93a,art93b,Miyoshi99}. Indeed, the difference is
compounded on top of another factor $\sim3$ decrease from models that
use the original torque cutoff function without the buffer region. 

Recent work by \citet{DHK02} has also studied the migration rates for
a range of planets from a few earth masses up to Jovian mass objects.
They also find that the migration rates from their simulations fall a
factor of several smaller in magnitude than the predictions of
theory. Given that the magnitude discrepancy has been found by
separate workers, using completely independent numerical codes, we
would speculate that it is a real effect. Regardless of the origin,
the conclusions reached in \jone\ only become stronger if the torques
are instead much larger than in our simulations. 

\citet{DHK02} briefly speculate that the reason for the discrepancy
is the existence of positive corotation torques on the planet, which
act to reduce the net migration rate. The results of our work would
seem to refute this supposition because we have separated out the
contributions due both to LR and CR torques and found that the torque
magnitude discrepancy exists in the LR torques independent of the CR
torques. The comparison between our work and theirs is particularly
interesting because the magnitude discrepancy extends to very small
planet masses, for which the system is certainly in the linear regime.
A linearized, numerical study using an iterative techniques
\citep{KP93} has studied the effects of both LR and CR torques, while
avoiding some of the approximations made in the purely analytic work.
In contrast to our results and \citet{DHK02}, they report good
agreement between analytic theory and their numerical model for LR
torques, especially for patterns for which the torque cutoff function
does not play a role. Their net LR torques were accurate to $\lesssim
50$\%, as opposed to the factor 3--6 discrepancy that we find.

Comparison of the three results is somewhat puzzling because all three
(\citet{KP93}, \citet{DHK02} and our study) consider two dimensional
systems and include gravitational softening (though with varying
amounts) and would therefore seem to solve the same physical problem
and be expected to yield the same results. It is interesting to
speculate that the differences between theory and simulation found by
us and \citet{DHK02} are due to the finite width of the resonances
and/or to resonance mixing, which causes one resonance and another to
partially cancel each other's torques, as discussed in MVS. Such
resonance mixing would be still more efficient if the corotation
resonances (not considered in the MVS analysis) participated in the
effect, since the inter-resonance distance is smaller and the overlap
is greater. No proof of this supposition can be found in our
simulations however.

It is also interesting to speculate that any resonance mixing phenomena
that are present may be due to a breakdown of some of the most basic
assumptions of resonances themselves. In simplest picture of a
resonant interaction of any kind, a particle is bound and oscillates
in a parabolic potential, driven by a small external force.
Applied to a star/planet/disk system, the (now approximately)
parabolic potential is a sum of several terms, including a stellar
potential and an centrifugal `effective potential' term that accounts
for the conservation of angular momentum. In cases with
\mplan=$0.1-1$\mj\  as in our study and many others, the driving term
(interaction with the planet) will no longer be small, especially in
the case where the planet and disk matter are within a few Hill radii
of each other. Indeed, the original potential may be perturbed so
strongly that it may no longer be parabolic at all. Such a condition
is certainly true any time the resonance position falls closer than
one Hill radius from the planet, where disk matter may no longer even
remain bound in the unperturbed potential, and may be true in more
limited circumstances at much larger distances. 

In addition to the LR torques, we also find significant torque
contributions due to corotation resonances.  A partial sum of the
torques from these patterns ($m\le15$), is smaller in magnitude than
the net LR torque but is opposite in sign (positive) and large enough
to decrease the total net torque on the planet by 20-30\%, below that
due to the LRs alone. Our results are in qualitative agreement with
\citet{KP93}, who found CR torques that were opposite in sign from the
LR torques for small $m$. Their results produced LR and CR torques
with comparable magnitude for a uniform surface density disk, but in a
disk where the gradient of $\Sigma/B$ was zero, produced no CR
torques, as expected. While our models begin with the same initial
condition, they quickly evolve away from it and large gradients
develop, allowing torques to develop.

If, as we have concluded, a large fraction of the gravitational torque
on the planet is exerted within one or a few Hill radii of the planet,
then it will be critical to resolve the three dimensional mass
distribution there in order to understand the global problem of the
planet's migration through the disk. Two distinct problems are
associated with this region. First, with the possible exception of the 
recent \citet{DHK02} work in 2d, no present models have resolved
the region within a few Hill radii of the planet well enough to
determine the gravitational torques with sufficient precision to make
an accurate determination of the migration rate. The full, three
dimensional mass distribution must be specified in order for such a
determination to be made. This description will require not only high
spatial resolution, but also a better thermodynamic model, since
energy release during the mass accretion will affect the thermodynamic
structure near the planet, changing the mass distribution.

Second, no current models have made an accurate determination of the
mass accretion rate onto the planet during the \tone\ migration era.
This is important for the migration because it determines how rapidly
the planet can grow large enough to open a gap via dynamical
processes. The mass accretion rate will be better known when models
that resolve the flow pattern and the thermodynamic state of the
region within a few Hill radii of the planet are developed. Improving
the description of the flow pattern will help to determine the
angular momentum input into the envelope and the likelihood of
rotational instabilities as it grows. Improving the thermodynamic
description of the gas will help to determine the mass accretion rate
into the envelope itself, and the frequency of those instabilities if
indeed they can grow. Work is now underway that will begin to address
both of these questions.

\acknowledgements
We would like to thank Willy Kley and Pawel Ciecielag for many
productive conversations during the evolution of this work. Bill Ward,
Pawel Artymowicz and Doug Lin each provided very helpful discussions
about the details of the torque calculations and Adri Olde Dalhuis
pointed out the work of \citet{Olvers}, which improved the accuracy of
our calculation of the Laplace coefficients. AFN thanks the UK
Astrophysical Fluids Facility (UKAFF) for support during the last
months during which this manuscript was prepared. Some of the
computations reported here were performed using the UK Astrophysical
Fluids Facility (UKAFF).

\singlespace
\begin{deluxetable}{lcccccc}
\tablewidth{0pt}
\tablecaption{\label{tab:table-acc} Initial Parameters For Simulations}
\tablehead{
\colhead{Name}  & \colhead{Resolution} & \colhead{Disk Mass} &
\colhead{Planet Mass}  & \colhead{Softening} & \colhead{Accretion} &
\colhead{Duration} 
\\
\colhead{}  & \colhead{($r\times\theta$)} & \colhead{$M_{\odot}$} &
\colhead{$M_{J}$}  &  \colhead{$\epsilon$} & \colhead{Factor, f} &
\colhead{(yr)} }

\startdata
acc0 & 128$\times$224 & 0.05 & 0.30     &  1.0\phn & 1$\times10^{0}$   &    1800 \\
acc1 & 128$\times$224 & 0.05 & 0.30     &  1.0\phn & 1$\times10^{-1}$  &    1800 \\
acc2 & 128$\times$224 & 0.05 & 0.30     &  1.0\phn & 1$\times10^{-2}$  &    1800 \\
acc3 & 128$\times$224 & 0.05 & 0.30     &  1.0\phn & 1$\times10^{-3}$  &    1800 \\
acc4 & 128$\times$224 & 0.05 & 0.30     &  1.0\phn & 1$\times10^{-4}$  &    1800 \\
gap0\tablenotemark{1} 
     & 128$\times$224 & 0.05 & 0.30     &  1.0\phn & 1$\times10^{0}$   &    1800 \\
gap1\tablenotemark{1} 
     & 128$\times$224 & 0.05 & 0.30     &  1.0\phn & 1$\times10^{-1}$  &    1800 \\
gap2\tablenotemark{1} 
     & 128$\times$224 & 0.05 & 0.30     &  1.0\phn & 1$\times10^{-2}$  &    1800 \\
gap3\tablenotemark{1} 
     & 128$\times$224 & 0.05 & 0.30     &  1.0\phn & 1$\times10^{-3}$  &    1800 \\
gap4\tablenotemark{1} 
     & 128$\times$224 & 0.05 & 0.30     &  1.0\phn & 1$\times10^{-4}$  &    1800 \\
mas3 & 128$\times$224 & 0.05 & 0.30     &  1.0\phn & 0.0               &    3000 \\
Sof7 & 256$\times$448 & 0.05 & 0.30     &  2.0\phn & 0.0               &    2400 \\
nosg\tablenotemark{2} 
     & 128$\times$224 & 0.05 & 0.30     &  1.0\phn & 0.0               &    1200 \\
Nosg\tablenotemark{2} 
     & 256$\times$448 & 0.05 & 0.30     &  2.0\phn & 0.0               & \phn500 \\
\enddata
\tablenotetext{1}{The simulations {\it gap0}--{\it gap4} are identical
to {\it acc0}--{\it acc4} except that they begin with a gap formed by
evolving the system for 3000~yr with the planet's migration and accretion
suppressed.} 
\tablenotetext{2}{The simulations {\it nosg} and {\it Nosg} do not include self 
gravity in the disk, but are otherwise identical to simulations {\it mas3} 
and {\it Sof7}, respectively. } 

\end{deluxetable} 

\doublespace

\singlespace

\clearpage

\clearpage

\clearpage

\clearpage

\clearpage

\clearpage

\clearpage

\clearpage

\clearpage

\clearpage

\clearpage

\clearpage

\clearpage

\clearpage

\clearpage

\clearpage

\clearpage

\begin{figure}
\psfig{file=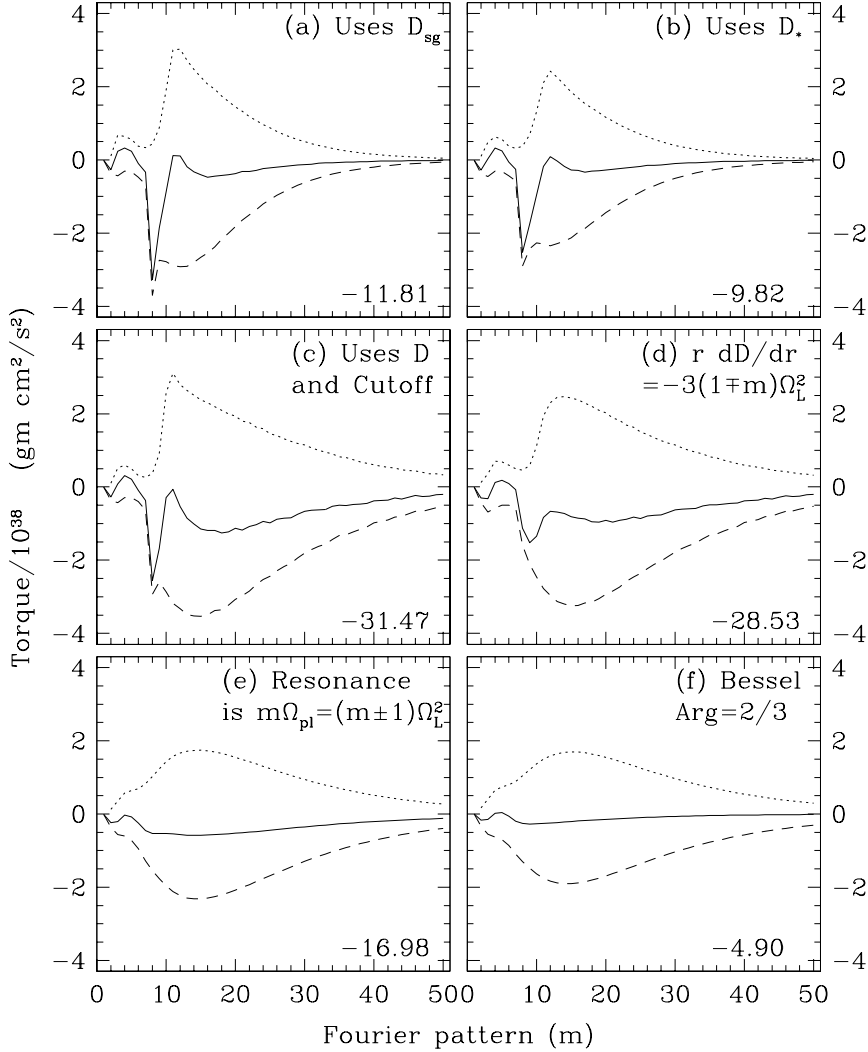,height=6.0in,rheight=5.65in}
\caption[Analytic Torque under different assumptions]
{\label{fig:torq-mod}
Analytic torques evaluated under various approximations to equation
\ref{eq:LR-flux}, using azimuth averaged quantities derived from the
high resolution prototype at $t=300$~yr. Successive panels (a)-(c)
investigate various approximations made to the theory, while panels
(c)-(f) investigate common approximations made to obtain the resonant
denominator and the resonance positions. Panel (a) is a duplication of
the analytic result shown in figure \ref{fig:mtorq-LR} for which the
torque is calculated using the exact resonance locations, determined
from $D_{sg}$, the generalized Laplace coefficients and the
numerically determined resonant denominator, $r_L dD_{sg}/dr$. Panel
(b) shows the same assumptions but with $D_*$ replacing $D_{sg}$.
Panel (c) replaces the combination of $D_*$ and generalized Laplace
coefficient with the combination of standard Laplace coefficients
(i.e. zero softening), resonant denominator, $D$ and the torque cutoff
function, equation \ref{eq:cutoff}. Panel (d) Uses the approximate
resonant denominator $r dD/dr|_{r_L}=-3(1\mp m)\Omega_L^2$ rather the
numerically obtained values, while retaining the numerically (from the
zeros of $D$) obtained resonance positions. Panel (e) uses the same
approximate resonant denominator, but now approximates the resonance
positions to be exact integer ratios of the planet's orbit frequency.
Panel (f) evaluates Bessel functions with an argument of 2/3 but uses
the exact frequency ratio to define $r dD/dr|_{r_L}$. Solid, dotted
and dashed curves are defined as in figures \ref{fig:mtorq-LR} and
\ref{fig:mtorq-LR-nosg}, above, as are the net torques shown in the
lower right corner of each panel. } 
\end{figure}

\clearpage

\begin{figure}
\psfig{file=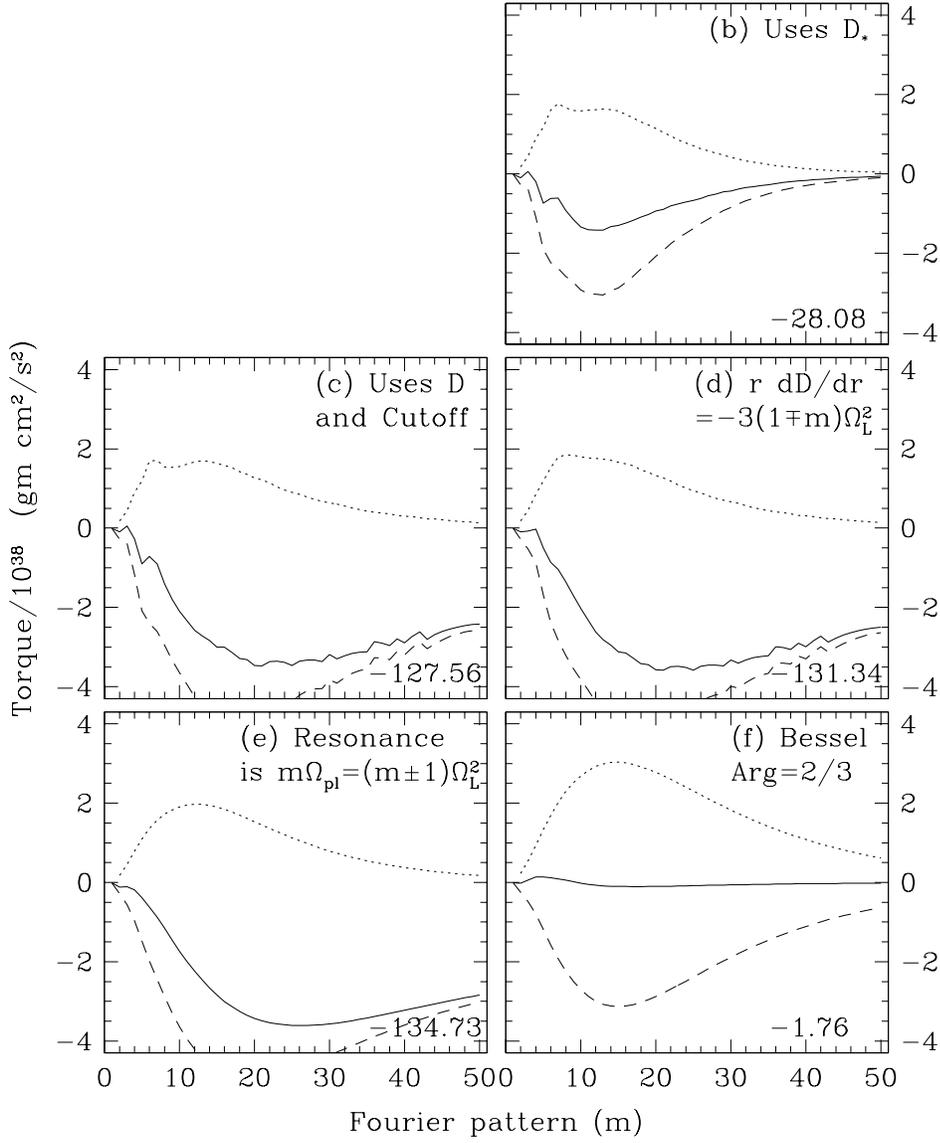,height=6.5in,rheight=6.5in}
\caption[Analytic Torque under different assumptions]
{\label{fig:torq-mod-nosg}
Same as figure \ref{fig:torq-mod}, but for the non-self gravitating
version of the high resolution prototype. Note that the (a) panel is
suppressed because by assumption self gravity is not present in this
version. }
\end{figure}

\begin{figure}
\psfig{file=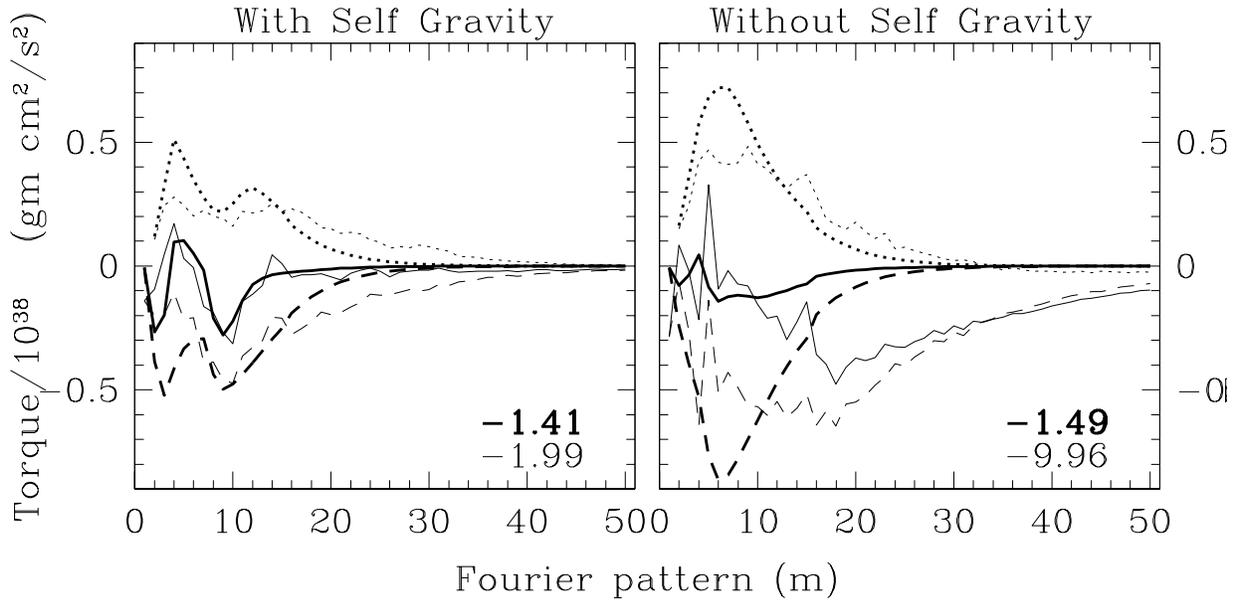,height=8.5in,rheight=4.5in}
\caption[Analytic Torque under different softening assumptions]
{\label{fig:torq-soft}
`Best' comparison of the simulation and analytically derived torques,
with and without disk self gravity. The heavy curves denote the
torques from the analytic calculation, while the light curves are for
the simulation. Solid, dotted and dashed curves defined as in figures
\ref{fig:mtorq-LR} and \ref{fig:mtorq-LR-nosg}, above, as
are the values for the net torques in the lower right corner of each
panel. In the generalized Laplace coefficient, we use a softening of
2.5$\times$ that in the simulation. We also use the approximation that
$|r dD_{sg}/dr|_{r_L}=-3(1\mp m)\Omega_{r_L}^2$, for $m\le15$.} 
\end{figure}

\notetoeditor{It is greatly to be desired that figures
\ref{fig:torq-mod} and \ref{fig:torq-mod-nosg} are placed on facing
pages}

\clearpage

\end{document}